\documentclass[conference]{IEEEtran}
\IEEEoverridecommandlockouts

\usepackage{multirow}
\usepackage{caption}

\usepackage{tikz}
\usepackage{amsmath}
\usepackage{mathtools}
\usepackage{amsthm}
\usepackage{subfiles}
\usepackage{subcaption}
\usepackage{pgfplots}
\usepackage{pgfplotstable}
\usepackage{booktabs}
\usepackage{diagbox}
\usepackage{multirow}
\usepackage{makecell}
\usepackage{xurl}
\usepackage[most]{tcolorbox}
\usepackage{lipsum}
\usepackage{hhline}
\usepackage{xspace}

\usepackage[T1]{fontenc}
\usepackage[frozencache,cachedir=.]{minted}

\usepackage[linesnumbered,ruled,noend]{algorithm2e}
\usepackage{algorithmic}
\usepackage{cleveref}

\usepackage{cite}
\usepackage{amsmath,amssymb,amsfonts}
\usepackage{algorithmic}
\usepackage{graphicx}
\usepackage{textcomp}
\usepackage[table]{xcolor}
\def\BibTeX{{\rm B\kern-.05em{\sc i\kern-.025em b}\kern-.08em
    T\kern-.1667em\lower.7ex\hbox{E}\kern-.125emX}}

\usetikzlibrary{calc}
\usetikzlibrary{fit}
\usetikzlibrary{patterns}
\usetikzlibrary{decorations.pathreplacing}
\usetikzlibrary{arrows, calc, arrows.meta, shapes, positioning}

\newcommand{\system}{{\sc MojoFrame}\xspace}
\newcommand{\systembf}{{\bf MojoFrame}\xspace}
\newcommand{\pandas}{{\sc Pandas}\xspace}
\newcommand{\modin}{{\sc Modin}\xspace}

\newcommand{\polars}{{\sc Polars}\xspace}

\definecolor{OursColor}{HTML}{C62E2E}
\definecolor{PandasColor}{HTML}{0081a7}
\definecolor{ModinColor}{HTML}{f4a261}
\definecolor{PolarsColor}{HTML}{6D6D6D}
\definecolor{Lightgrey}{HTML}{dadada}

\begin{document}

\title{MojoFrame: Dataframe Library in Mojo Language}

\author{\IEEEauthorblockN{Shengya Huang, Zhaoheng Li, Derek Warner, Yongjoo Park}
\IEEEauthorblockA{University of Illinois Urbana-Champaign\\
\{shengya4, zl20, derekw3, yongjoo\}@illinois.edu}
}

\maketitle

\begin{abstract}

Mojo is an emerging programming language built on MLIR (Multi-Level Intermediate Representation) and supports JIT (Just-in-Time) compilation. 
It enables transparent hardware-specific optimizations (e.g., for CPUs and GPUs),
    while allowing users to express their logic
        using Python-like user-friendly syntax.
Mojo has demonstrated strong performance on tensor operations;
however, its capabilities for relational operations (e.g., filtering, join, and group-by aggregation) common in data science workflows, remain unexplored.
To date, no dataframe implementation exists in the Mojo ecosystem.

In this paper, we introduce the first Mojo-native dataframe library, 
    called \system, 
    that supports core relational operations and user-defined functions (UDFs).
\system is built on top of Mojo's tensor to achieve fast operations on numeric columns, 
while utilizing a cardinality-aware approach 
to effectively integrate non-numeric columns for flexible data representation.
To achieve high efficiency,
\system takes significantly different approaches than 
    existing libraries.
We show that \system supports all operations for TPC-H queries and a selection of TPC-DS queries with promising performance, achieving up to $4.60\times$ speedup versus existing dataframe libraries in other programming languages.
Nevertheless,
there remain optimization opportunities for \system (and the Mojo language), particularly in in-memory data representation and dictionary operations.

\end{abstract}

\section{Introduction}
\label{sec:intro}

Relational operations such as filter, join, and group-by aggregation are integral to data science tasks such as data analysis~\cite{xin2021enhancing, vartak2015seedb, kraska2021northstar}, data cleaning~\cite{chu2016data, de2013introduction}, feature engineering~\cite{gao2021scaling, dong2018feature}, and data visualization~\cite{fang2025enhancing, fang2025large, fang2023learning}. These operations are typically performed using dataframes—a flexible, table-like data structure widely adopted in data science. Compared to traditional database tables, dataframes offer greater flexibility in schema evolution and heterogeneous data type support~\cite{petersohn2020towards, wu2020dataframe, petersohn2021dataframe}.
Dataframe libraries supporting these relational operations are present in many popular programming languages employed in data science. For example, Pandas~\cite{pandas} and Modin~\cite{petersohn2021flexible} for Python, Polars~\cite{polars} for Rust, \texttt{frames} in R~\cite{rdataframe}, and Spark's dataframe~\cite{sparkframe} in Scala.
These libraries feature distinct pros and cons attributed to their native language. For instance, Pandas and Modin support flexible data types, but can be slow for user-defined functions (UDFs)~\cite{pandasudf}. Polars supports lazy executions for multi-operation queries~\cite{pllazyframe} but does not support user-defined objects~\cite{polarsdtypes}.

\paragraph{Dataframes in Mojo: Promising Alternative}
Mojo is a recent programming language with flexible, Python-like syntax specifically designed for data science while addressing many of the aforementioned shortcomings. Key capabilities include JIT~\cite{kulkarni2011jit} with MLIR~\cite{lattner2020mlir} for increased runtime efficiency, native CPU-GPU programming, and optimized tensor operations~\cite{mojo}. Mojo has been benchmarked on data science tasks like tensor and model operations, outperforming both Python~\cite{mojopythonbenchmark} and Rust~\cite{mojorustbenchmark}. Yet, performing relational operations in Mojo is currently unexplored due to the absence of a native dataframe~\cite{mojodataframe}, which we introduce in this paper (\cref{fig:mojo_intro}). We hypothesize from existing benchmarking results that such a dataframe library (which we call \system) would be a promising alternative to existing solutions, notably Pandas and Polars, achieving higher efficiency (especially on UDFs) versus the former, while being more scalable across different hardware targets versus the latter.



\paragraph{Challenges for MojoFrame}
Implementing a Mojo dataframe library that is both expressive and efficient in performing relational operations is challenging. 

First, Mojo is optimized for tensor operations on numeric data. Although this enables high-performance numeric computations through optimizations like SIMD, these techniques do not directly extend to non-numeric types such as strings, which are common in data science workflows. Efficiently supporting these heterogeneous data types while maintaining performance requires careful design decisions.

Second, Mojo is still relatively new and currently lacks mature implementations of certain data structures (e.g., handling mutable pointers~\cite{mojopointer} in dictionaries~\cite{mojodictionary}). These are commonly used by established dataframe implementations in other languages (e.g., Python's Pandas) for efficiently performing relational operations using conventional algorithms (e.g., hash join). Hence, we need to design novel adaptations.
\begin{figure}[t]
\tikzset{
codenode/.style={minimum width=16mm,
    font=\footnotesize\ttfamily,
},
cellnode/.style={
    font=\small\sffamily,
    anchor=south west,inner ysep=0,
},
technode/.style={
    draw=black,
    align=center,
    font=\footnotesize\sffamily,
}
}

\centering
\begin{tikzpicture}

\def\g{0.6}

\node[codenode, align=center] (C1) at (-3,0) {\texttt{FILTER} \\ \texttt{JOIN} \\ \texttt{GROUPBY} \\ \texttt{...}};

\node[draw=black,dashed,thick,fit={(C1)}, inner sep=0mm] (B) {};
\node[anchor=south, align=center, font=\small\bfseries] at (B.north) 
    {Relational\\ Operations};

\node[draw=black,dashed,thick,anchor=north west,
    minimum width=45mm,minimum height=18mm] 
    (B2) at ($(B.north east)+(\g,0.45)$) {};
\node[anchor=south,font=\small\bfseries] 
    at ($(B2.north)+(0,-0.08)$)
    {Mojo Programming Language};
\node[anchor=north west,align=center,technode, minimum height=7.5mm,minimum width = 10mm] 
    (T1) at ($(B2.north west)+(0.1,-0.1)$)
    {JIT};
\node[anchor=north west,technode, minimum height=7.5mm,minimum width = 10mm] 
    (T2) at ($(T1.south west)+(0, -0.1)$)
    {\baselineskip=0pt MLIR};
\node[anchor=north west,technode, minimum height=16mm,minimum width=32mm, inner sep=0mm] 
    (T3) at ($(T1.north east)+(0.1, 0)$)
    {\baselineskip=0pt Our Mojo-native \\ dataframe (\systembf)};

\node[draw=black,dashed,thick,anchor=north west,
    minimum width=12mm,minimum height=18mm,align=center,
    font=\small\sffamily] 
    (B3) at ($(B2.north east)+(\g,0)$) {
        CPU \\
        GPU \\
        TPU \\
        ...
    };
\node[anchor=south,font=\small\bfseries] 
    at ($(B3.north)+(0,-0.02)$)
    {Compute};

\draw[-latex, ultra thick] 
    ($(B.east)+(0.05,0.225)$) -- ($(B2.west)+(-0.05,-0)$);
\draw[latex-, ultra thick] 
    ($(B.east)+(0.05,-0.175)$) -- ($(B2.west)+(-0.05,-0.4)$);
\draw[-latex, ultra thick] 
    ($(B2.east)+(0.05,0.2)$) -- ($(B3.west)+(-0.05,0.2)$);
\draw[latex-, ultra thick] 
    ($(B2.east)+(0.05,-0.2)$) -- ($(B3.west)+(-0.05,-0.2)$);
    
\end{tikzpicture}
\caption{\systembf (ours) is the first Mojo-native dataframe library.
Mojo is a new language with JIT, MLIR, designed for compatibility with heterogeneous hardware (CPU/GPU).
}
\label{fig:mojo_intro}
\end{figure}
\paragraph{Our Approach}

We design \system by balancing three pillars: efficiency through optimized tensor operations, flexibility in handling heterogeneous data types, and scalability across diverse hardware backends. Our implementation utilizes Mojo's native tensors for numeric data and operations whenever appropriate. Then, we derive alternative techniques for tasks where tensor-based approaches are not directly applicable, such as high-cardinality string operations and complex joins.

First, for data representation, numeric columns are directly stored in a tensor. For non-numeric columns, we derive a cardinality-aware approach. Low-cardinality columns are mapped to numeric values and integrated into the tensor. High-cardinality columns are offloaded to separate data structures (e.g., lists). Our dataframe enables this offloading in a column and row order-preserving manner, with a decoupled physical and logical layouts enabled via indexers.

Second, for relational operations such as group-by aggregation, we develop implementations adapted to Mojo's currently limited ecosystem. Rather than relying on dictionaries with mutable keys---which Mojo does not yet adequately support~\cite{mojodictionary}---we devise tuple-based hashing combined with list indexing to track distinct grouping keys. Where suitable, we leverage vectorization and parallelization to maximize hardware potential.


\paragraph{Comparison against Other Methods for GPU-Based Relational Operations}
\system differs from existing GPU-accelerated frameworks in approach and focus. GPU dataframe libraries like cuDF~\cite{treinen2008description} and cuPy~\cite{nishino2017cupy} aim to port explicit Pandas and NumPy functions to GPUs for improved computational efficiency. \system instead focuses on native dataframe abstraction within Mojo's portable programming model, i.e., \textit{how to} represent and perform dataframe operations, where GPU compatibility emerges from hardware-agnostic compilation as opposed to back-end specific code. GPU database systems like BlazingSQL~\cite{ocsa2019sql} and Crystal~\cite{crystal} orthogonally target SQL query processing on rigid schemas, addressing different use cases than \system's flexible, exploratory dataframe workflows.

\paragraph{Contributions} According to our motivations (\cref{sec:background}), we implement \system to achieve the following:
\begin{itemize}
  \item \textbf{Universal Representation.} We introduce \system's representation and how it supports the variety of datatypes commonly used in data science (\cref{sec:framework}).
  \item \textbf{Relational Operations Support.} We describe our implementations to support filtering, group-by aggregation, and joins in \system (\cref{sec:operations}).
  \item \textbf{TPC-H Benchmark.} We show \system's support for all existing 22 TPC-H queries and select TPC-DS queries, and benchmark its performance versus alternative dataframes (\cref{sec:exp}). 
\end{itemize}

\setminted{
  fontsize=\scriptsize, 
  numbersep=3pt, 
  frame=single,
  framesep=1mm, 
  xleftmargin=1mm, 
  xrightmargin=1mm, 
  breaklines=true,
  breakanywhere=true,
}

\section{Background}
\label{sec:background}

This section describes the Mojo language (\cref{sec:background_mojo}), existing dataframe implementations in other languages (\cref{sec:background_others}), and finally, how a Mojo-based dataframe can benefit existing data science pipelines (\cref{sec:background_gap}).

\subsection{What is Mojo?}
\label{sec:background_mojo}

This section describes Mojo's key characteristics that enable its high performance and adaptability across various data science tasks.

\paragraph{Mojo's Just-in Time (JIT) Compilation} Mojo is a compiled language that supports JIT. Like other JIT-based languages such as Java, Mojo's JIT compilation allows generation of optimized machine code specific to the hardware it is running on, achieving better performance on a variety of operations present in data science tasks, such as tensor operations and complex UDFs, versus interpreted languages such as Python and R~\cite{mojopythonbenchmark} (\cref{fig:mojo_udf}). While Mojo's JIT compilation potentially incurs latency, Mojo code can also be compiled ahead-of-time~\cite{whymojo} (e.g., for repeated use during recurring operations~\cite{li2023s}). We empirically verify that this (dataset-size agnostic) latency is often negligible compared to the time saved from faster data loading, processing, etc, especially on tasks using larger dataset scales (\cref{fig:exp_compile}).

\paragraph{Multi-level Intermediate Representation (MLIR)} Mojo is built natively on MLIR~\cite{lattner2020mlir}, a recent compiler infrastructure designed for domain-specific optimizations. When executing code for TPC-H query processing~\cite{barata2015overview}, the compiler has the ability to progressively generate lower-level intermediate representations at runtime, applying domain-specific optimizations (e.g., data reading/tensor computations) where necessary. In contrast, non-MLIR frameworks (e.g., TensorFlow Graph~\cite{tfgraph}) require coordinating multiple specialized compilers for different hardware backends (e.g., GPU, TPU) to achieve progressive lowering. MLIR's single extensible framework simplifies maintaining data science workloads as both hardware and algorithms evolve~\cite{whymojo}.

\subsection{Existing Dataframes for Data Science}
\label{sec:background_others}

In this section, we overview the large variety of dataframe libraries available to other programming languages, and accordingly hypothesize potential benefits that \system can bring over them.









\paragraph{Dataframes in Python}
Python features many popular dataframe libraries such as Pandas~\cite{pandas}, Modin~\cite{modin}, and Dask~\cite{dask}.
Pandas, built on NumPy arrays, supports heterogeneous data types across columns with vectorized operations, though it can also accommodate mixed types within a column using object dtype.
Modin is a drop-in Pandas replacement that is able to parallelize operations such as transpose and pivot.
Dask is a Pandas-based distributed dataframe library for parallel operations on large datasets.

Since each library targets different use cases (e.g., Modin for parallelized operations, Dask for distributed workloads), users often switch between libraries to optimize different stages in data science pipelines, causing data conversion and workflow overhead~\cite{baziotis2024dias}. \system eliminates this fragmentation. As a Mojo-native library, it benefits directly from compiler and hardware optimization~\cite{mojo}. It provides a unified, low-overhead alternative to multiple Python libraries.\footnote{%
To clarify, Mojo does not currently support distributed computing natively.
}

\paragraph{Dataframes in Other Compiled Languages}
There exists dataframe libraries in other compiled languages such as Rust (Polars~\cite{polars}) and Julia (JuliaFrame~\cite{juliaframe}).
While these dataframes also natively support optimizations like parallelism, their more complex syntax creates a steeper learning curve. With Mojo's Python-like syntax, \system can bring the performance benefits and still preserve ease of use.

\paragraph{GPU Dataframes}
Specialized GPU dataframes like cdDF~\cite{treinen2008description} and cuPy~\cite{nishino2017cupy} enable accelerated CPU-GPU computations, with additional research on optimizations for data placement~\cite{yogatama2024scaling} and JIT compilation~\cite{yogatama2023accelerating}. However, these libraries target CUDA exclusively~\cite{farber2011cuda} and require substantial code modifications when using alternative backends such as MPS~\cite{pytorchmps} or Intel~\cite{pytorchintel}. In contrast, Mojo provides transparent integration with multiple GPUs (currently NVIDIA, AMD, and Apple~\cite{mojo}). Hence, \system is a potentially more generalizable alternative; code runs on any available hardware without platform-specific modifications (\cref{fig:mojo_gpu}).\footnote{To achieve unified scalability across hardware backends, implementations must follow Mojo's portable programming model~\cite{mojocustomops} (the Tensor abstraction used in \system is scalable). A more detailed tutorial can be found on the GPU programming fundamentals page~\cite{mojotutorial}.}


\begin{figure}[t]
\centering

\begin{subfigure}[b]{\linewidth}
    \begin{minted}[fontsize=\scriptsize, xleftmargin=1em]{cuda}
template <typename Op>
__global__ void reduce(float *in, float *out, float init, Op op) {
    // NVIDIA-specific
    __shared__ float sdata[BLOCK_SIZE]; 
    int tid = threadIdx.x;
    sdata[tid] = op(init, in[blockIdx.x * blockDim.x + tid]);
    __syncthreads(); // NVIDIA-specific

    for (unsigned int s = blockDim.x / 2; s > 0; s >>= 1) {
        if (tid < s) sdata[tid] = op(sdata[tid], sdata[tid + s]);
        __syncthreads();
    }
    if (tid == 0) out[blockIdx.x] = sdata[0];
}
    \end{minted}
    \caption{Reduction kernel in CUDA (C++)}
\end{subfigure}

\vspace{2mm}

\begin{subfigure}[b]{\linewidth}
    \begin{minted}[fontsize=\scriptsize, xleftmargin=1em]{mojo}
fn reduce[op: fn(Float32, Float32) -> Float32](in: Tensor, out: Tensor, init: Float32):
    # Portable across vendors
    var sdata = tensor_builder().shared().alloc() 
    var tid = thread_idx.x
    sdata[tid] = op(init, input[block_idx.x * BLOCK_SIZE + tid])
    barrier() # Portable across vendors

    var s = BLOCK_SIZE // 2
    while s > 0:
        if tid < s: sdata[tid] = op(sdata[tid], sdata[tid + s])
        barrier()
        s //= 2

    if tid == 0: output[block_idx.x] = sdata[0]
    \end{minted}
    \caption{Reduction kernel in Mojo}
\end{subfigure}

\caption{Mojo adopts a portable, vendor-independent GPU programming model~\cite{mojo}, enabling execution on heterogeneous hardware (NVIDIA, AMD, Apple) without backend-specific code. This allows parts of relational operations suitable for numeric kernels (e.g., aggregations) to be portable.}
\label{fig:mojo_gpu}
\end{figure}

\subsection{Mojo-Native Data Science Pipelines}
\label{sec:background_gap}
This section describes the potential benefits that implementing \system brings to end-to-end data science pipelines.

\paragraph{Python Data Science Pipelines}
Data science tasks in Python, such as data cleaning, feature engineering, and visualization, require various libraries. For example, data scientists may find themselves loading data into a Pandas dataframe~\cite{pandas} for data analysis, converting the dataframe into on-GPU PyTorch~\cite{pytorch} or Tensorflow~\cite{tftensor} tensors for training models, converting to Scipy~\cite{scipy} to sparsify output tensors for row/column operations, and finally, plotting with Matplotlib~\cite{matplotlib} or Seaborn~\cite{seaborn}.
This fragmented ecosystem incurs inefficiencies due to the requirement for data conversions (e.g., from Pandas to PyTorch) and additional management overhead from maintaining compatible library versions.

\paragraph{Mojo Data Science Pipelines}
Mojo is designed to natively support the complete data science lifecycle~\cite{mojopromise}, for which it currently includes built-in tools (i.e., similar to Python's standard library~\cite{pythonstandard}) for various general (e.g., model training) and more specialized (e.g., CV, NLP) data science tasks~\cite{mojodatascience}. Despite the current lack of a Mojo-native dataframe, users can work around this limitation in Mojo by importing Python's Pandas, as Mojo supports Python libraries through an integrated CPython runtime~\cite {cpython}. However, such an approach leads to a similar data conversion inefficiency as observed in Python data science pipelines, requiring conversions between Mojo-native types (e.g., \texttt{Int32}) used by Mojo libraries and generic \textit{Python Objects} used by imported Python libraries. Therefore, implementing \system and completing a Mojo-native data science pipeline is important both for performance benefits and ease of use: users only need to maintain a single cohesive codebase with no additional external dependencies.

\section{MojoFrame: Data Representation}
\label{sec:framework}
\subfile{figures/framework}

This section describes our implementation of \system. While Mojo's design naturally enables efficient tensor computation, naively translating existing dataframe libraries from other programming languages is insufficient. For example, a translation of the dynamically-typed Pandas in Mojo will fail to leverage Mojo's static typing and hardware acceleration for performance. Hence, we take an approach that idiomatically maps heterogeneous columns into Mojo's static type system, as depicted in \Cref{fig:system_overview}.



\paragraph{Data Loading}
\system supports loading dataframes stored in common file formats supported by existing dataframe implementations (CSV, Parquet~\cite{parquet}, ORC~\cite{orc}, Arrow~\cite{arrow}, etc., \cref{sec:discussion}). Once loaded, \system organizes the dataframe columns based on their column types (e.g., numeric vs. non-numeric, high vs. low cardinality) into respective components within \system (described shortly).

\paragraph{Tensor}
\system's tensor stores all numeric columns, for example, the integer column \texttt{int1} and float column \texttt{float1} in \cref{fig:system_overview}. It can also retrieve the element indices of low-cardinality non-numeric columns, which are mapped into the tensor.

\paragraph{Low-cardinality Non-numeric Columns}
\system maintains non-numeric columns with cardinality below a user-defined threshold by mapping distinct elements to dense integer identifiers. This approach is applied to columns such as \texttt{str1} and \texttt{str2}. By storing these dense integers within the tensor, \system achieves efficient, numeric-based processing for relational operations like filtering and group-by aggregation (\cref{sec:operations}). 

\paragraph{High-cardinality Non-numeric Columns}
\system offloads non-numeric columns with cardinality above the user-defined threshold (i.e., \textit{high cardinality}), for example, \texttt{str3} and \texttt{str4} into lists separate from the tensor. This approach notably differs from the multi-array \texttt{BlockManager} approach of Pandas dataframe~\cite{pandas}, which distributes columns strictly by type (e.g., all integer columns in an \texttt{array[int]}, all string columns in an \texttt{array[str]}); this effectively offloads all columns into separate arrays. \system maps low-cardinality columns into its tensor when doing so enables higher operation efficiency via tensor operations (TPC-H Q19, \cref{sec:exp_e2e}).





\paragraph{Column Names}
\system stores column names like other dataframe implementations such as Pandas and Polars. This allows \system to support column-based operations which refer to column names, e.g., \texttt{df['min'].fillna()}.

\paragraph{Row and Column Indexers}
\system's row and column indexers control the logical layout independently of the physical layout of columns in the tensor and offloaded high-cardinality non-numeric lists. For example, given the row and column indexers depicted in \cref{fig:system_overview}, the ordering of columns in the \system is \texttt{str3}, \texttt{str1}, \texttt{int1}, \texttt{float1}, \texttt{str2}, and \texttt{str4}. This approach allows \system to logically interleave numeric and non-numeric columns regardless of their positioning in the tensor/lists. It also efficiently supports relational operations that potentially alter row and/or column orders, such as joins and groupbys, as only the indexers require updates while the physical data layout remains unchanged (\cref{sec:operations}).

\section{MojoFrame Operations}

\label{sec:operations}
This section presents our approach to supporting relational operations in \system. Since Mojo differs significantly from other programming languages that host dataframe libraries, many techniques used in existing dataframes for relational operations are not effectively translatable. We describe our approaches to filtering in \cref{sec:operations_filter}, group-by aggregation in \cref{sec:operations_groupby}, and joins in \cref{sec:operation_join}.



\subsection{Filtering}
\label{sec:operations_filter}
This section describes our approach to support filtering in \system.
Simple filters such as equality, greater-than, and less-than can be implemented with boolean indexing in existing dataframe libraries such as Pandas and Polars (e.g., via a mask \texttt{df['A'] < 5}), which are then executed with vectorized instructions. However, dataframe filtering often requires specifying custom logic (e.g., regular expressions) beyond these simple comparisons. Such user-defined functions (UDFs) are not typically vectorized and instead executed ``row-by-agonizing-row''~\cite{fritchey2014row} in existing dataframes~\cite{polarsapply, pandasapply}.



\paragraph{Filtering in Existing Dataframe Libraries}
Existing libraries, notably Pandas~\cite{pandasapply} and Polars~\cite{polarsapply}, enable filtering with complex UDFs through the \texttt{df.apply()} interface, which allows users to define and apply boolean-returning lambda functions for filtering (e.g., \texttt{lambda x: re.search('\%special\%request\%', x)} in TPC-H Q13).
However, like UDFs in database transactions~\cite{duta2020functional, zhang2023automated}, the lambda functions passed to \texttt{apply()} can be stateful (i.e., the result of one row depends on the application results of prior rows). Consequently, given Python's interpreted environment, the lambda function (even when not stateful) must be executed row-by-row without parallelization in Pandas and Modin.
This can potentially be alleviated with Numba's JIT compilation~\cite{pandasapply}, which unfortunately does not support some commonly-used data science operations (e.g., \texttt{argsort}~
\cite{numbaargsort}).
Polars, while natively implemented in Rust, still fails to take advantage of Rust's compilation for parallelized UDF execution due to it internally offloading lambda function applications to Python for expressiveness~\cite{polarsapply}.

\begin{figure}[t]
\begin{subfigure}[b]{\linewidth}
    \begin{minted}[fontsize=\scriptsize]{python}
df.apply(lambda x, cmp: math.sin(2 * math.pi * x) > math.cos(2 * math.pi * cmp))  # Cyclical feature engineering for timestamps
    \end{minted}
    \caption{Pandas \texttt{apply()}}
\end{subfigure}
\begin{subfigure}[b]{\linewidth}
\vspace{2mm}
    \begin{minted}[fontsize=\scriptsize]{python}
fn evaluate(self, x: SIMD[DType.float64, 1], cmp: SIMD[DType.float64, 1]):
    var p = 2 * math.pi * x
    var q = 2 * math.pi * cmp
    return True if math.sin(p) > math.cos(q) else False
    \end{minted}
    \caption{\system trait}
\end{subfigure}

    \caption{\system's parallelized filtering with stateless lambda functions vs. Pandas' sequential filtering with \texttt{apply}.}
    \label{fig:mojo_filter}
\end{figure}
\paragraph{\system's Approach}
We aim to support parallelized filtering with stateless lambda functions in \system. To accomplish this, we introduce a trait-based filtering mechanism that allows users to define expressive, generic filter conditions which inherit from a set of stateless, extensible, and JIT-optimizable base operations (e.g., equality, greater/less than, assignment, mathematic and string operations, \cref{fig:mojo_filter}).
This allows the Mojo compiler to parallelize filter operations defined through inheriting traits, as they are guaranteed to be stateless (lambda functions), and produce optimization passes more effectively (compared to ad-hoc inline definition). This approach is more efficient than handling equivalent UDFs with \texttt{apply()} in existing dataframe implementations (empirically verified in \cref{sec:exp_operator}).

\subsection{Group-by Aggregation}
\label{sec:operations_groupby}
This section describes our approach to support group-by aggregation in \system.
Efficient multi-column group-by aggregation presents algorithmic and system-level challenges: algorithmically, multi-column group-by aggregation requires the creation of \textit{composite keys} (i.e., a combination of the $k$ keys in each row for a $k$-column group-by) and finding distinct keys. The number of possible composite keys grows exponentially with the number of grouping columns. Hence, system-wise, multi-column aggregation demands efficient composite key management and mapping for finding distinct keys.

\paragraph{Group-by Aggregation in Existing Dataframe Libraries}
Existing dataframe libraries such as Pandas employ a sparse-to-dense, incremental composite key creation and mapping strategy (summarized in \cref{alg:pandas_groupby}). 
As these dataframes store data in column-major data storage for efficient memory access~\cite{pandas, modin, polars}, each of the $k$ columns involved in the group-by is processed sequentially for incremental composite key and hash generation (line 4).
For a dataframe with $n$ rows, $n$ composite keys (internally stored as conceptual lists) and accompanying incremental hashes are maintained. For each column, the sparse-to-dense step is first applied to map unique elements to integer identifiers (i.e., like \system's mapping for low-cardinality non-numeric columns, \cref{sec:framework}) (line 5). Then, these integer identifiers are incrementally collected into the $n$ composite keys (line 7), while the $n$ hashes are incrementally updated via a vectorization-friendly arithmetic combination~\cite{pandashash} (line 8). Finally, the $n$ composite keys are inserted into a dictionary with the $n$ hashes to find distinct groups (line 9).



\begin{algorithm}[t]
\caption{\small{Pandas Column Order Group-By} \label{alg:pandas_groupby}}
\SetKw{KwIn}{Input:}
\SetKw{KwOut}{Output:}
\small
\KwIn{$n$-row dataframe $df$, group-by columns $g_1,..., g_k$}
\KwOut{$n' \leq n$ unique comp. keys $\{df[i,1]...df[i,k]\}$, $1 \leq i \leq n'$}
Initialize $n$ empty composite keys $C_i = []$, $1 \leq i \leq n$; \\
Initialize $n$ empty hashes $H_i = 0$, $1 \leq i \leq n$; \\
Initialize $k$ unique element-to-identifier mappings $M_{g_c} = \{\}$, $1 \leq c \leq k$; \\

\For{each column $df[:,g_c]$}{
  Compute element-to-identifier mapping: $M_{g_c}: df[:,g_c] \rightarrow \mathbb{N}$\\
  \For{each element $df[j, g_c]$ in column $df[:,g_c]$}{ 
    $C_j.append(M_{g_c}(df[j, g_c]))$;\\
    $H_j.update(M_{g_c}(df[j, g_c]))$;\\
  }
}
Insert $(C_i, H_i), 1 \leq i \leq n$ into dictionary to find unique composite keys;\\
\textbf{Return} $\{df[i,1],..., df[i,k]\}$, $1 \leq i \leq n'$.
\end{algorithm}

\paragraph{Incremental Hashing in Mojo}
An analogous approach to Pandas' incremental hashing for \system would be to similarly maintain a list of $n$ composite keys (stored as lists) and incremental hashes (for a $n$-row dataframe) and process the group-by columns in per-column order: elements are collected into the $n$ composite key-lists and hashed using a generalizable hash function (due to offloaded non-numeric columns possibly appearing in the group-by, \cref{sec:framework}) such as xxhash~\cite{xxhash} to incrementally update the $n$ hashes. Finally, the composite keys would be inserted along with the hashes into Mojo's dictionary. Unfortunately, this approach is currently inefficient because Mojo's dictionary does not support mutable keys (i.e., the composite key-lists) by reference. Inserting a mutable class instance into Mojo's dictionary triggers a \textit{deep copy}~\cite{mojodictionary}, incurring significant time/memory overhead (PandasMojo, \cref{fig:mojo_groupby_bench}).\footnote{There is an in-progress update to Mojo's dictionary class to optimize the copying of mutable keys.\cite{mojodictionarybug}.}

\begin{algorithm}[t]
\caption{\small{MojoFrame Transposed Group-By} \label{alg:mojo_groupby}}
\SetKw{KwIn}{Input:}
\SetKw{KwOut}{Output:}
\small
\KwIn{$n$-row dataframe $df$, groupby columns $g_1,...,g_k$}\\
\KwOut{$n'\!\leq\! n$ unique comp. keys $\{df[i,1]...df[i,k]\}$, $1\!\leq\!i\!\leq\! n'$ }\\
Initialize composite key array $C_i$, $1 \leq i \leq n$; \\
Initialize hash array $H_i$, $1 \leq i \leq n$; \\
Transpose $dft = df.T$;\\
\For{each column $dft[:,i]$ in transposed dataframe}{
  $C_i \leftarrow tuple(dft[g_1,i],...,dft[g_k,i])$;\\
  $H_i \leftarrow hash(dft[g_1,i],...,dft[g_k,i])$;\\
}
Insert $(C_i, H_i), 1 \leq i \leq n$ into dictionary to find unique composite keys;\\
\textbf{Return} $\{df[i,1],...,df[i,k]\}$, $1 \leq i \leq n'$.
\end{algorithm}

\paragraph{\system's Approach (\cref{alg:mojo_groupby})}
To leverage efficient immutable keys and avoid strided memory access across columns, \system first \textit{transposes} the grouping columns into a contiguous row-major layout (line 5). This enables the single-pass construction of $n$ immutable tuples as composite keys (line 7) alongside $n$ non-incremental hashes (line 8). Then, we insert these $n$ tuples along with the non-incremental hashes into Mojo's dictionary, which are not duplicated due to the tuples being immutable (line 9). We empirically justify and verify this approach in \cref{sec:exp_operator}.

\subsection{Join}
\label{sec:operation_join}
This section describes our approach to support inner join in \system.\footnote{We defer supporting other join types not present in the TPC-H benchmark~\cite{tpch}, such as cross join and various outer joins, to future work.}
Similar to group-by, joining large dataframes presents the algorithmic challenge of efficiently identifying equivalent keys~\cite{gao2021scaling}, but across two relations.



\begin{algorithm}[t]
\caption{\small{MojoFrame (and Pandas) Factorize-then-Join}} \label{alg:mojo_join}
\SetKw{KwIn}{Input:}
\SetKw{KwOut}{Output:}
\small
\KwIn{Dataframes $df_x, df_y$, join columns $j_1,..., j_k$}\\
\KwOut{Joined dataframe $df_{\text{join}}$}\\
Initialize mappings set $F = \{\}$ for non-numeric columns; \\
\For{each non-numeric join column $c$ in $j_1,..., j_k$}{
  Compute mapping $f_c: \text{unique}(df_x[c] \cup df_y[c]) \rightarrow \mathbb{N}$;\\
  Compute mapped columns: $df_x[c] \leftarrow f_c(df_x[c])$, $df_y[c] \leftarrow f_c(df_y[c])$;\\
}
Compute left/right row indexers via hash probe on factorized columns;\\
Materialize $df_{\text{join}}$ using indexers (parallelized vector gather);\\
\textbf{Return} $df_{\text{join}}$.
\end{algorithm}

\paragraph{Adopting Pandas' Join Algorithm to \system}
Existing dataframe libraries, like Pandas, implement a hash join derivative to perform inner join. Unlike traditional DBMSs that hash values in the join columns directly during the build and probe phases, Pandas adds a pre-processing step: non-numeric join columns are first \textit{factorized} into a shared integer space~\cite{pandasfactorize} similar to \system's mapping of low-cardinality non-numeric columns to integer identifiers (\cref{sec:framework}). Then, these identifiers are processed following the standard hash join algorithm~\cite{pandasjoin} (summarized in \cref{alg:mojo_join}). 
This strategy is advantageous because performing hash join on the factorized integers (via sequential integer arrays) is more memory-efficient than direct hash computation and collision detection on non-numeric columns~\cite{koutris2025quest, pandasfactorize, barber2014memory}.
We find this approach translates well to the Mojo language and hence is suitable for our dataframe. Unlike Pandas, \system's factorize-then-hash-join is parallelized via compilation, enabling faster join execution (empirically verified in \cref{sec:exp_operator}).

\section{Discussion}
\label{sec:discussion}
\paragraph{Coding with MojoFrame} We design MojoFrame's programming interface following the typical per-operation, chained method call style of popular dataframe libraries~\cite{keshk2023method}. \Cref{fig:mojo_example} depicts examples of TPC-H Q16 in \pandas and \system, which involves filtering, join, and group-by aggregation. There are one-to-one correspondences between many MojoFrame and Pandas operations (e.g., \texttt{merge} vs. \texttt{inner\_join}), which we design as such to facilitate easier usage by users already familiar with dataframe operations. One notable difference is \system's trait-based filtering requiring users to separately define filter masks versus Pandas' direct application of boolean masks. However, this design choice is crucial for enabling \system's optimization of (stateless) filter applications via compilation (\cref{sec:operations_filter}).
\begin{figure}[t]
\begin{subfigure}[b]{\linewidth}
    \begin{minted}[fontsize=\scriptsize]{python}
# Data read
df_supplier = pd.read_csv('supplier.dat')
df_part = pd.read_csv('part.dat')
df_ps = pd.read_csv('partsupp.dat')

# Filter
df_part_filtered = df_part[
    (df_part['p_brand'] != 'Brand#45') &
    (~df_part['p_type'].str.startswith('MEDIUM POLISHED')) &
    (df_part['p_size'].isin([49, 14, 23, 45, 19, 3, 36, 9]))
]
df_supplier_filtered = df_supplier[df_supplier['s_comment'].apply(lambda comment: not_string_exists_before(comment, "Customer", "Complaints"))]

# Join
joined_pss_df = pd.merge(df_partsupp, df_supplier_filtered, left_on='ps_suppkey', right_on='s_suppkey', how='inner')
joined_psp_df = pd.merge(joined_pss_df, df_part_filtered, left_on='ps_partkey', right_on='p_partkey', how='inner')

# Aggregation
result = joined_psp_df.groupby('p_size').agg(supplier_count=('ps_suppkey', 'nunique'))
result = result.sort_values(by=['supplier_count', 'p_size'])]
    \end{minted}
    \caption{Pandas TPC-H Q16 implementation}
    \vspace{2mm}
\end{subfigure}
\begin{subfigure}[b]{\linewidth}
    \begin{minted}[fontsize=\scriptsize]{python}
# Data read
var df_supplier = DataFrame('supplier.dat')
var df_part = DataFrame('part.dat')
var df_ps = DataFrame('partsupp.dat')

# Filter
var p_brand_mask = string_not_equal(p_brand, "Brand#45")
var p_type_mask = string_not_startwith(p_type, "MEDIUM POLISHED")
var p_size_mask = f64_IN(df_part["p_size"], [49, 14, 23, 45, 16, 3, 36, 9])
df_part.select_complex([p_brand_mask, p_type_mask, p_size_mask], "AND")
not_string_exists_before(df_supplier, s_comment, "Customer", "Complaints")

# Join
var joined_pss_df = inner_join(df_ps, df_supplier, "suppkey")
var joined_psp_df = inner_join(joined_pss_df, df_part, "partkey")

# Aggregation
joined_psp_df.groupby("p_size", "count_distinct", ["p_size", "suppkey"])
joined_psp_df.sort_by(["suppkey", "p_size"])
    \end{minted}
    \caption{\system TPC-H Q16 implementation}
\end{subfigure}

    \caption{TPC-H Q16 with \system and Pandas. \system follows the per-operation, chained method call programming style of popular dataframe libraries.}
    \label{fig:mojo_example}
\end{figure}
\paragraph{Efficient Data Loading} While \system is compatible with common data formats (\cref{sec:framework}), there currently lacks an optimized native CSV parser in Mojo. Thus, we implement a custom binary data adaptor resembling that of Polars~\cite{polarsparquet}. This enables projection pushdown (loading only necessary columns) and allows us to benchmark \system's native I/O and memory management capabilities. As shown in \cref{sec:exp_dataload}, this implementation is faster than runtime data parsing and full table loading, which occur in existing dataframes during CSV parsing.
\paragraph{Extensibility of \system}
Mojo is a language under active development, which is reflected in some of \system's custom implementations to address temporary limitations present in the language. However, many of the identified limitations, especially those affecting popular open-source Mojo libraries, have been quickly fixed by members of the Modular team. For example, optimizations for string construction~\cite{mojostringopt} and an in-progress fix for \system's mutable dictionary keys~\cite{mojodictionarybug}. \system has recently attracted interest from members of the Mojo community~\cite{mojoframediscord}. Hence, we expect \system's performance to be further improved as the ecosystem matures, aided by the ongoing development of a Mojo-based centralized package manager~\cite{mojoroadmap} like Python's PyPi, and the open-sourcing of Mojo's compiler.
\section{Experiments}
\label{sec:exp}

In this section, we empirically study the effectiveness of \system. Although our current evaluation focuses on dataframe libraries to align closely with MojoFrame’s intended usage scenario, interactive dataframe analytics, we recognize the value of future detailed performance comparisons with SQL-first engines employing LLVM-based compilation. We organize our evaluation into two categories:

\noindent\textbf{Query Execution \& Scalability:}
\begin{enumerate}
    \item \textbf{Analytical Query Processing Time:} \system achieves faster execution runtimes for UDF-heavy queries and low-cardinality group-by aggregation compared to existing dataframe libraries (\cref{sec:exp_e2e}).
    \item \textbf{Scalability of \systembf to Large Datasets:} \system exhibits linear scalability with respect to dataset size, a characteristic typical of parallelized dataframes (\cref{sec:exp_large}).
    \item \textbf{Parallelism of \systembf:} \system achieves speedup with increasing core counts compared to existing single-threaded dataframe implementations, and we further benchmark against parallelized dataframes (\cref{sec:exp_cores}).
\end{enumerate}
\noindent
\textbf{System Microbenchmarks:}
\begin{enumerate}
    \item \textbf{Relational Operator Performance:} We study \system's performance on individual filter, join, and group-by aggregation operations versus other dataframe libraries (\cref{sec:exp_operator}).
    \item \textbf{Microbenchmark on Compilation Time:} We study the compilation overhead of \system incurred by Mojo's JIT compilation, and show that it is both largely agnostic to query complexity and negligible versus query runtime at large dataset scales (\cref{sec:exp_compilation}).
    \item \textbf{Microbenchmark on Data Loading Time:} We investigate \system's data loading times in the Mojo programming language versus data loading times of alternative dataframe implementations in Python (\cref{sec:exp_dataload}).
    \item \textbf{Microbenchmark on In-Memory Data Size:} We investigate \system's in-memory table sizes versus those of alternative dataframe implementations (\cref{sec:exp_datasize}).
\end{enumerate}

\subsection{Experiment Setup}
\label{sec:exp_setup}
We use the table generator and queries included in the \textit{TPC-H}~\cite{tpch} decision support benchmark in our experiments.
We generate TPC-H datasets from 4 distinct scale factors (1, 3, 10, 100); the scale factor determines the total size in GB of the tables in the generated dataset. All data tables are stored in the CSV format.

\paragraph{Workload} We use all 22 TPC-H queries and 5 selected TPC-DS queries covering diverse relational operations for our workload. As the queries are written in SQL, we translate the queries into equivalent code for each dataframe library (e.g., SQL's \texttt{GROUPBY} into Pandas' \texttt{agg()}) for evaluation. To focus on evaluating numerical computation efficiency, our translations explicitly load only the columns necessary for the core relational logic.

\paragraph{Methods}
We evaluate \system by comparing it to the following established dataframe libraries commonly used in data science:
\begin{enumerate}
    \item \textbf{Pandas~\cite{pandas}:} We perform all operations in Pandas with default function arguments (e.g., no Numba JIT~\cite{numbaargsort}).
    \item \textbf{Modin~\cite{modin}:} A drop-in Pandas alternative that parallelizes common operations (e.g., group-by, transpose). We similarly use default arguments for all operations.
    \item \textbf{Polars~\cite{polars}:} A dataframe library natively implemented in Rust. We use this library in Python via its Python bindings, with lazy execution disabled for fair comparison on operator performance.
\end{enumerate}
For \system, we compile our SQL-to-Mojo translated queries ahead-of-time for execution. However, we also study setups where we perform JIT compilation as part of query execution (\cref{sec:exp_compilation}).
For non-numeric column storage, we treat columns with cardinality greater than 50\% of the row count as high-cardinality (stored without mapping), while mapping other string columns (\cref{sec:framework}). As a result, all comment-type columns (e.g., \texttt{partsupp}'s \texttt{comment}) are stored without mapping.\footnote{We defer incorporating techniques for automatically determining this threshold to future work.}

\paragraph{Environment} 
All experiments are performed on a Standard E16ads v6 Azure machine with 16 vCPUs (AMD EPYC 9004 Genoa) and 128GB RAM. Input data is read from a local NVMe SSD disk with 165.44 MB/s read speed.\footnote{Measured with \texttt{hdparm -Tt /dev/sda.}, sanity checked with ~\cite{azurebenchmark}} We use 8 cores for most of our experiments. However, we also study setups with fewer cores in \cref{sec:exp_cores}. 

\paragraph{Implementation}
We implement \system natively in Mojo following the data structure described in \cref{sec:framework} and running relational operations as described in \cref{sec:operations}. We manually implement some functions (e.g., substring matching with regexes, TPC-H Q13) not directly translatable from Python to Mojo (due to lack of libraries, e.g., regex~\cite{pythonregex}) required for some operations in the TPC-H queries.
\system supports all 22 translated TPC-H queries with these additional function implementations.


\paragraph{Time measurement}
We pre-load all datasets into memory to mimic interactive data science scenarios. We measure the \textit{query execution runtime} as the time from invoking the query on the in-memory tables to observing results. Specific to \system, we also study the \textit{compile time} as time incurred by Mojo's JIT compilation in cases where ahead-of-time compilation is not used in \cref{sec:exp_compilation}. For data loading, we study the \textit{data read time} for reading relevant input tables into memory in \cref{sec:exp_dataload}. We run each query/operation 5 times and report the average. We clear the page cache between runs.

\paragraph{Reproducibility}
Our \system implementation and TPC-H, TPC-DS queries translated into Mojo and other benchmarked languages can be found in our Github repository.\footnote{https://github.com/illinoisdata/MojoFrame}





\begin{figure*}[t]
\usetikzlibrary{patterns}
\begin{subfigure}[b]{\linewidth}
\centering
\begin{tikzpicture}

\pgfplotstableread[col sep=comma,]{
name
Q1
Q2
Q3
Q4
Q5
Q6
Q7
Q8
Q9
Q10
Q11
Q12
Q13
Q14
Q15
Q16
Q17
Q18
Q19
Q20
Q21
Q22
}\datatable

\begin{axis}[
    ybar,
    clip=false,
    xtick={1, 2, 3, 4, 5, 6, 7, 8, 9, 10, 11, 12, 13, 14, 15, 16, 17, 18, 19, 20, 21, 22},
    xticklabels from table={\datatable}{name},
                 x tick label style={anchor=center, yshift = 0ex, font=\scriptsize},
    xtick style ={draw=none},
    xlabel style={yshift = 2.5ex, font=\footnotesize},
    ylabel style={yshift = -1.5ex, font=\scriptsize},
    width=180mm,
    height=35mm,
    bar width=0.75mm,
    ymin=0,
    ymax=5.1,
    axis y line*=none,
    axis x line*=none,
    ytick={0, 1, 2, 3, 4, 5},
    yticklabels={0\%, 100\%, 200\%, 300\%, 400\%, 500\%},
    xmin=0.5,
    xmax = 22.5,
    ymajorgrids,
    tick label style={font=\footnotesize},
    legend style={
        font=\footnotesize,
        /tikz/every even column/.append style={column sep=0.2cm},
        legend columns = 4,
        at={(0.46,1.25)},
        anchor=center,
        /tikz/every even column/.append style={column sep=4pt},
    },
    ylabel={Time \% vs. Pandas},
    area legend
    ]

\addplot[black, fill=PandasColor]
table[x=x,y=y] {
x y
1 1 
2 1 
3 1
4 1
5 1
6 1 
7 1
8 1
9 1
10 1
11 1
12 1
13 1
14 1
15 1
16 1
17 1
18 1
19 1
20 1
21 1
22 1
};
\addlegendentry{\pandas}

\addplot[black, fill=ModinColor]
table[x=x,y=y] {
x y
1 1.248954012
2 1.830975543
3 3.74067352
4 1.570224695
5 2.320476691
6 4.000880009
7 1.006356047
8 1.908507163
9 0.5044074712
10 1.021653863
11 1.922767058
12 3.562538929
13 4.932020965
14 5.5
15 2.08549083
16 2.936438692
17 0.8809079487
18 1.080503724
19 1.84764319
20 1.459144835
21 0.9557613486
22 1.73767488
};
\addlegendentry{\modin}

\addplot[black, fill=PolarsColor]
table[x=x,y=y] {
x y
1 0.2977079392
2 1.309159821
3 0.9626825621
4 0.7180586397
5 1.890652488
6 0.9045434913
7 0.09023883914
8 0.317164839
9 0.2819967368
10 1.763330964
11 0.4511581654
12 5.12089374
13 0.7914247844
14 0.4484131102
15 0.276737244
16 0.2256383129
17 1.00176617
18 0.1688731819
19 0.182223022
20 0.2236942479
21 0.6424060448
22 0.2340631755
};
\addlegendentry{\polars}

\addplot[black, fill=OursColor]
table[x=x,y=y] {
x y
1 0.653849992
2 0.3592993202
3 1.042895581
4 0.2877093976
5 0.6201482214
6 0.6491045718
7 0.3248547655
8 0.4042055046
9 0.06917894572
10 0.3681352192
11 0.3972654877
12 1.408267453
13 0.1722474271
14 0.5447652905
15 0.4335076236
16 0.3092557719
17 0.5621983387
18 0.651550292
19 0.2799425639
20 0.9406882808
21 0.7630922698
22 0.4760560382
};
\addlegendentry{\system \textbf{(Ours)}}

    
        \draw[fill=white,draw=white] (axis cs: 13.7,5.1) -- (axis cs: 14.1,5.2) -- (axis cs: 14.1,5.1) -- (axis cs: 13.7,5.0) -- cycle;
    \draw[draw=black] (axis cs: 14.1,5.2) -- (axis cs: 13.7,5.1);
    \draw[draw=black] (axis cs: 14.1,5.1) -- (axis cs: 13.7,5.0);
    
    \node[anchor=west] at (axis cs: 14.1,5.1) {\footnotesize 51.9$\times$};

\end{axis}
    
\end{tikzpicture}
\label{fig:experiment_checkout_undo}
\end{subfigure}

\caption{\system's normalized query execution times (w.r.t. Pandas) on the 22 TPC-H queries versus alternative dataframes. \system is up to $4.60\times$ faster than the next best alternative on UDF-heavy queries (e.g., Q13) and low-cardinality aggregation (e.g., Q9), but falls short on high-cardinality aggregation (e.g., Q18).}
\label{fig:exp_e2e}
\end{figure*}
\begin{figure*}[t]

\centering

\begin{subfigure}[b]{0.24\linewidth}
\begin{tikzpicture}

\begin{axis}[
    xtick=data,
    clip=false,
    width=45mm,
    height=32mm,
    ymin=0.1,
    ymax=1000,
    log origin = infty,
    xmode=log,
    ymode=log,
    axis y line*=none,
    axis x line*=none,
    xtick={1,3,10, 100},
    xticklabel style = {align=center},
    xticklabels={1GB,3GB,10GB,100GB},
    ytick={0.1, 1, 10, 100, 1000},
    yticklabels={0.1, 1, 10, 100, 1000},
    xlabel=Dataset Scale,
        xlabel style={yshift = 1.5ex},
    ylabel style={yshift=-2.5ex},
    xmin = 1,
    xmax = 100,
    tick label style={font=\footnotesize},
    legend style={
        at={(1.3,1.2)},anchor=south west,column sep=2pt,
        draw=black,fill=white,
        /tikz/every even column/.append style={column sep=5pt},
        font=\footnotesize,
    },
    legend cell align={left},
    legend columns=4,
    label style={font=\footnotesize},
    ylabel={Runtime (s)},
    ymajorgrids,
]

\addplot[PandasColor, thick, mark = *, mark size=1pt]
table[x=x,y=y] {
x y
1 0.9062833786010742
3 4.57417726516723
10 12 
100 140.2
};
\addplot[ModinColor, thick, mark = *, mark size=1pt]
table[x=x,y=y] {
x y
1 0.8282999992370605
3 2.3112263679504395
10 10.260703802108765 
100 603.67
};
\addplot[PolarsColor, thick, mark = *, mark size=1pt]
table[x=x,y=y] {
x y
1 0.41107294199900934
3 1.3644605609988503 
10 6.037358044999564
100 140.25
};
\addplot[OursColor, thick, mark = *, mark size=1pt]
table[x=x,y=y] {
x y
1 0.437902282
3 1.371610164
10 5.449402129
100 33.481
};
\addlegendentry{\pandas}
\addlegendentry{\modin}
\addlegendentry{\polars}
\addlegendentry{\system \textbf{(Ours)}}

\end{axis}
\end{tikzpicture}
\caption{Q9}
\label{fig:scalability_query1}
\end{subfigure}
\begin{subfigure}[b]{0.24\linewidth}
\begin{tikzpicture}

\begin{axis}[
    xtick=data,
    width=45mm,
    height=32mm,
    ymin=0.1,
    ymax=1000,
    log origin = infty,
    xmode=log,
    ymode=log,
    axis y line*=none,
    axis x line*=none,
    xtick={1,3,10, 100},
    xticklabel style = {align=center},
    xticklabels={1GB,3GB,10GB,100GB},
    ytick={0.1, 1, 10, 100, 1000},
    yticklabels={0.1, 1, 10, 100, 1000},
    xlabel=Dataset Scale,
        xlabel style={yshift = 1.5ex},
    ylabel style={yshift=-2.5ex},
    xmin = 1,
    xmax = 100,
    tick label style={font=\footnotesize},
    legend style={
    inner sep=1.5pt,
        at={(-0.2,1.1)},anchor=south west,column sep=2pt,
        draw=black,fill=white,
        /tikz/every even column/.append style={column sep=5pt},
        font=\footnotesize,
    },
    legend cell align={left},
    legend columns=4,
    label style={font=\footnotesize},
    ylabel={Runtime (s)},
    ymajorgrids,
]

\addplot[PandasColor, thick, mark = *, mark size=1pt]
table[x=x,y=y] {
x y
1 0.7005734443664551
3 2.08040380477905
10 6.89662544558
100 84.558
};
\addplot[ModinColor, thick, mark = *, mark size=1pt]
table[x=x,y=y] {
x y
1 1.752357006072998
3 5.422057390213013
10 19.289525032043457
100 192.155
};
\addplot[PolarsColor, thick, mark = *, mark size=1pt]
table[x=x,y=y] {
x y
1 0.5356600939994678
3 1.588843954999902 
10 5.175659036998695
100 54.362
};
\addplot[OursColor, thick, mark = *, mark size=1pt]
table[x=x,y=y] {
x y
1 0.154680049
3 0.480634631
10 1.751486674
100 18.168
};

\end{axis}
\end{tikzpicture}
\caption{Q13}
\end{subfigure}
\begin{subfigure}[b]{0.24\linewidth}
\begin{tikzpicture}

\begin{axis}[
    xtick=data,
    width=45mm,
    height=32mm,
    ymin=0.01,
    ymax=100,
    log origin = infty,
    xmode=log,
    ymode=log,
    axis y line*=none,
    axis x line*=none,
    xtick={1,3,10, 100},
    xticklabel style = {align=center},
    xticklabels={1GB,3GB,10GB,100GB},
    ytick={0.01, 0.1,1, 10, 100},
    yticklabels={0.01, 0.1,1, 10, 100},
    xlabel=Dataset Scale,
        xlabel style={yshift = 1.5ex},
    ylabel style={yshift=-2.5ex},
    xmin = 1,
    xmax = 100,
    tick label style={font=\footnotesize},
    legend style={
    inner sep=1.5pt,
        at={(-0.2,1.1)},anchor=south west,column sep=2pt,
        draw=black,fill=white,
        /tikz/every even column/.append style={column sep=5pt},
        font=\footnotesize,
    },
    legend cell align={left},
    legend columns=4,
    label style={font=\footnotesize},
    ylabel={Runtime (s)},
    ymajorgrids,
]

\addplot[PandasColor, thick, mark = *, mark size=1pt]
table[x=x,y=y] {
x y
1 0.39849019050598145
3 1.15989470481872
10 3.81713184518
100 46.01
};
\addplot[ModinColor, thick, mark = *, mark size=1pt]
table[x=x,y=y] {
x y
1 0.49619579315185547
3 1.037055492401123
10 2.866445541381836
100 26.50
};
\addplot[PolarsColor, thick, mark = *, mark size=1pt]
table[x=x,y=y] {
x y
1 0.0312539039987314
3 0.09780085400052485
10 0.3313377529993886
100 2.887
};
\addplot[OursColor, thick, mark = *, mark size=1pt]
table[x=x,y=y] {
x y
1 0.357682078
3 1.136701364
10 4.097509433
100 12.724
};
\end{axis}
\end{tikzpicture}
\caption{Q19}
\end{subfigure}
\begin{subfigure}[b]{0.24\linewidth}
\begin{tikzpicture}

\begin{axis}[
    xtick=data,
    clip=false,
    width=45mm,
    height=32mm,
    ymin=0.1,
    ymax=1000,
    log origin = infty,
    xmode=log,
    ymode=log,
    axis y line*=none,
    axis x line*=none,
    xtick={1,3,10, 100},
    xticklabel style = {align=center},
    xticklabels={1GB,3GB,10GB,100GB},
    ytick={0.1, 1, 10, 100, 1000},
    yticklabels={0.1, 1, 10, 100, 1000},
    xlabel=Dataset Scale,
        xlabel style={yshift = 1.5ex},
    ylabel style={yshift=-2.5ex},
    xmin = 1,
    xmax = 100,
    tick label style={font=\footnotesize},
    legend style={
    inner sep=1.5pt,
        at={(-0.2,1.1)},anchor=south west,column sep=2pt,
        draw=black,fill=white,
        /tikz/every even column/.append style={column sep=5pt},
        font=\footnotesize,
    },
    legend cell align={left},
    legend columns=4,
    label style={font=\footnotesize},
    ylabel={Runtime (s)},
    ymajorgrids,
]

\addplot[PandasColor, thick, mark = *, mark size=1pt]
table[x=x,y=y] {
x y
1 0.8506381511688232
3 4.01271820068359
10 16.2205781936645
100 208.827
};
\addplot[ModinColor, thick, mark = *, mark size=1pt]
table[x=x,y=y] {
x y
1 1.171569585800171 
3 3.513143301010132
10 13.814715385437012
100 172.22
};
\addplot[PolarsColor, thick, mark = *, mark size=1pt]
table[x=x,y=y] {
x y
1 0.2621433439999237 
3 0.8654497819989047
10 3.050837787999626
100 34.130
};
\addplot[OursColor, thick, mark = *, mark size=1pt]
table[x=x,y=y] {
x y
1 1.329673669
2 4.476284373
10 20.125163353
100 136.702
};

\end{axis}
\end{tikzpicture}
\caption{Q21}
\end{subfigure}

\caption{\system's query processing times versus baseline dataframe implementations on various dataset scales. \system exhibits linear scaling versus dataset scale like existing parallelized dataframe implementations (Polars, Modin).}
\label{fig:exp_scalability}
\end{figure*}
\begin{figure*}[t]

\centering

\begin{subfigure}[b]{0.24\linewidth}
\begin{tikzpicture}

\begin{axis}[
    xtick=data,
    width=45mm,
    height=32mm,
    ymin=0,
    ymax=50,
    axis y line*=none,
    axis x line*=none,
    xtick={1,2,3},
    xticklabel style = {align=center},
    xticklabels={2, 4, 8},
    ytick={0, 10, 20, 30, 40, 50},
    yticklabels={0, 10, 20, 30, 40, 50},
    xlabel=Number of Cores,
    xlabel style={yshift = 1.5ex},
    ylabel style={yshift=-4ex},
    xmin = 0.8,
    xmax = 3.2,
    tick label style={font=\footnotesize},
    legend style={
        at={(1.3,1.2)},anchor=south west,column sep=2pt,
        draw=black,fill=white,
        /tikz/every even column/.append style={column sep=5pt},
        font=\footnotesize,
    },
    legend cell align={left},
    legend columns=4,
    label style={font=\footnotesize},
    ylabel={Runtime (s)},
    ymajorgrids,
]

\addplot[PandasColor, thick, mark = *, mark size=1pt]
table[x=x,y=y] {
x y
1 43.2
2 43.2
3 43.2
};
\addplot[ModinColor, thick, mark = *, mark size=1pt]
table[x=x,y=y] {
x y
1 23.5146958827972
2 13.560675859451294
3 10.260703802108765 
};
\addplot[PolarsColor, thick, mark = *, mark size=1pt]
table[x=x,y=y] {
x y
1 13.146251559001
2 8.078691873000935 
3 6.037358044999564
};
\addplot[OursColor, thick, mark = *, mark size=1pt]
table[x=x,y=y] {
x y
1 6.375627871
2 5.644276198
3 5.447658624
};
\addlegendentry{\pandas}
\addlegendentry{\modin}
\addlegendentry{\polars}
\addlegendentry{\system \textbf{(Ours)}}

\end{axis}
\end{tikzpicture}
\caption{Q9}
\label{fig:scalability_query1}
\end{subfigure}
\begin{subfigure}[b]{0.24\linewidth}
\begin{tikzpicture}

\begin{axis}[
    xtick=data,
    width=45mm,
    height=32mm,
    ymin=0,
    ymax=40,
    axis y line*=none,
    axis x line*=none,
    xtick={1,2,3},
    xticklabel style = {align=center},
    xticklabels={2, 4, 8},
    ytick={0, 10, 20, 30, 40},
    yticklabels={0, 10, 20, 30, 40},
    xlabel=Number of Cores,
    xlabel style={yshift = 1.5ex},
    ylabel style={yshift=-4ex},
    xmin = 0.8,
    xmax = 3.2,
    tick label style={font=\footnotesize},
    legend style={
    inner sep=1.5pt,
        at={(-0.2,1.1)},anchor=south west,column sep=2pt,
        draw=black,fill=white,
        /tikz/every even column/.append style={column sep=5pt},
        font=\footnotesize,
    },
    legend cell align={left},
    legend columns=4,
    label style={font=\footnotesize},
    ylabel={Runtime (s)},
    ymajorgrids,
]

\addplot[PandasColor, thick, mark = *, mark size=1pt]
table[x=x,y=y] {
x y
1 6.89662544558
2 6.89662544558
3 6.89662544558
};
\addplot[ModinColor, thick, mark = *, mark size=1pt]
table[x=x,y=y] {
x y
1 35.6790120601654
2 24.1201753616333
3 19.289525032043457
};
\addplot[PolarsColor, thick, mark = *, mark size=1pt]
table[x=x,y=y] {
x y
1 5.791151498000545
2 5.44913562900183
3 5.323193524003727
};
\addplot[OursColor, thick, mark = *, mark size=1pt]
table[x=x,y=y] {
x y
1 2.328112821
2 1.985842444
3 1.740400196
};

\end{axis}
\end{tikzpicture}
\caption{Q13}
\end{subfigure}
\begin{subfigure}[b]{0.24\linewidth}
\begin{tikzpicture}

\begin{axis}[
    xtick=data,
    width=45mm,
    height=32mm,
    ymin=0,
    ymax=8,
    axis y line*=none,
    axis x line*=none,
    xtick={1,2,3},
    xticklabel style = {align=center},
    xticklabels={2, 4, 8},
    ytick={0, 2, 4, 6, 8},
    yticklabels={0, 2, 4, 6, 8},
    xlabel=Number of Cores,
    xlabel style={yshift = 1.5ex},
    ylabel style={yshift=-4ex},
    xmin = 0.8,
    xmax = 3.2,
    tick label style={font=\footnotesize},
    legend style={
    inner sep=1.5pt,
        at={(-0.2,1.1)},anchor=south west,column sep=2pt,
        draw=black,fill=white,
        /tikz/every even column/.append style={column sep=5pt},
        font=\footnotesize,
    },
    legend cell align={left},
    legend columns=4,
    label style={font=\footnotesize},
    ylabel={Runtime (s)},
    ymajorgrids,
]

\addplot[PandasColor, thick, mark = *, mark size=1pt]
table[x=x,y=y] {
x y
1 3.81713184518
2 3.81713184518
3 3.81713184518
};
\addplot[ModinColor, thick, mark = *, mark size=1pt]
table[x=x,y=y] {
x y
1 7.35333800315856
2 4.297097206115723
3 2.866445541381836
};
\addplot[PolarsColor, thick, mark = *, mark size=1pt]
table[x=x,y=y] {
x y
1 0.725219826003012
2 0.4438921949986252
3 0.3313377529993886
};
\addplot[OursColor, thick, mark = *, mark size=1pt]
table[x=x,y=y] {
x y
1 4.13
2 4.097509433
3 4.097509433
};

\end{axis}
\end{tikzpicture}
\caption{Q19}
\end{subfigure}
\begin{subfigure}[b]{0.24\linewidth}
\begin{tikzpicture}

\begin{axis}[
    xtick=data,
    width=45mm,
    height=32mm,
    ymin=0,
    ymax=21,
    axis y line*=none,
    axis x line*=none,
    xtick={1,2,3},
    xticklabel style = {align=center},
    xticklabels={2, 4, 8},
    ytick={0, 5, 10, 15, 20},
    yticklabels={0, 5, 10, 15, 20},
    xlabel=Number of Cores,
    xlabel style={yshift = 1.5ex},
    ylabel style={yshift=-4ex},
    xmin = 0.8,
    xmax = 3.2,
    tick label style={font=\footnotesize},
    legend style={
    inner sep=1.5pt,
        at={(-0.2,1.1)},anchor=south west,column sep=2pt,
        draw=black,fill=white,
        /tikz/every even column/.append style={column sep=5pt},
        font=\footnotesize,
    },
    legend cell align={left},
    legend columns=4,
    label style={font=\footnotesize},
    ylabel={Runtime (s)},
    ymajorgrids,
]

\addplot[PandasColor, thick, mark = *, mark size=1pt]
table[x=x,y=y] {
x y
1 16.1097326937
2 16.1097326937
3 16.1097326937
};
\addplot[ModinColor, thick, mark = *, mark size=1pt]
table[x=x,y=y] {
x y
1 15.5030016899108
2 14.621418237686157 
3 13.814715385437012
};
\addplot[PolarsColor, thick, mark = *, mark size=1pt]
table[x=x,y=y] {
x y
1 10.4201974809984
2 5.619599128003756
3 3.050837787999626
};
\addplot[OursColor, thick, mark = *, mark size=1pt]
table[x=x,y=y] {
x y
1 20.783759591
2 20.569721065
3 20.125163353
};

\end{axis}
\end{tikzpicture}
\caption{Q21}
\end{subfigure}

\caption{\system's query processing times versus baseline dataframe implementations on variable number of cores.}
\label{fig:exp_cores}
\end{figure*}

\subsection{Fast In-Memory Analytics}
\label{sec:exp_e2e}

This section evaluates \system's performance on typical relational operations in analytical queries. We measure \system's query execution times on all 22 TPC-H queries on the 10GB TPC-H dataset versus existing dataframe implementations, with all times normalized w.r.t. Pandas' runtime on the same query.

We report results in \cref{fig:exp_e2e}. \system exhibits highly competitive execution speeds, achieving faster execution times than Pandas, Modin, and Polars on 20, 22, and 10 out of the 22 queries, respectively.




\paragraph{Fast UDF Application}\system demonstrates a significant advantage on Q13, which contains a complex string filtering UDF (\cref{fig:mojo_udf}). It is $4.60\times$, $5.81\times$, and $28.6\times$ faster than Polars (next best alternative), Pandas, and Modin, respectively. This performance gain comes from \system's ability to compile stateless UDF logic via our trait-based system, enabling parallelized execution (\cref{sec:exp_operator}).


\paragraph{Group-By Aggregation Performance}
\system is $4.07\times$, $14.4\times$, and $7.28\times$ faster than Polars, Pandas, and Modin, respectively, on Q9, which contains a 2-column group-by aggregation applied on a large table (5 joins) with a small number of distinct groups. This highlights the efficiency of \system's transposed group-by strategy (\cref{sec:operations_groupby}), which maximizes memory locality for dense aggregations.
However, \system is slower than alternative dataframe implementations on queries involving high-cardinality aggregations (e.g., Q18). We further investigate this trend in \cref{sec:exp_operator} and discuss this finding in the limitations section.

\begin{figure}[t]
\usetikzlibrary{patterns}
\begin{subfigure}[b]{\linewidth}
\centering
\begin{tikzpicture}

\pgfplotstableread[col sep=comma,]{
name
Q3
Q6
Q7
Q9
Q96
}\datatable

\begin{axis}[
    ybar,
    clip=false,
    xtick={1, 2, 3, 4, 5},
    xticklabels from table={\datatable}{name},
                 x tick label style={anchor=center, yshift = 0ex, font=\scriptsize},
    xtick style ={draw=none},
    xlabel style={yshift = 2.5ex, font=\footnotesize},
    ylabel style={yshift = -1.5ex, font=\scriptsize},
    width=90mm,
    height=35mm,
    bar width=0.75mm,
    ymin=0,
    ymax=5.1,
    axis y line*=none,
    axis x line*=none,
    ytick={0, 1, 2, 3, 4, 5},
    yticklabels={0\%, 100\%, 200\%, 300\%, 400\%, 500\%},
    xmin=0.5,
    xmax = 5.5,
    ymajorgrids,
    tick label style={font=\footnotesize},
    legend style={
        font=\footnotesize,
        /tikz/every even column/.append style={column sep=0.2cm},
        legend image post style={xscale=0.6},
        legend columns = 4,
        at={(0.46,1.25)},
        anchor=center,
    },
    ylabel={Time \% vs. Pandas},
    area legend
    ]

\addplot[black, fill=PandasColor]
table[x=x,y=y] {
x y
1 1 
2 1 
3 1
4 1
5 1
};
\addlegendentry{\pandas}

\addplot[black, fill=ModinColor]
table[x=x,y=y] {
x y
1 0.7285018209
2 0.3701648628
3 1.011155025
4 4.572527275
5 0.7896723582
};
\addlegendentry{\modin}

\addplot[black, fill=PolarsColor]
table[x=x,y=y] {
x y
1 0.1440249705
2 0.07879732739
3 0.6783731485
4 0.1452261554
5 0.7494426544
};
\addlegendentry{\polars}

\addplot[black, fill=OursColor]
table[x=x,y=y] {
x y
1 0.3755043851
2 0.2761616517
3 0.383644497
4 0.376841593
5 0.4613226245
};
\addlegendentry{\system \textbf{(Ours)}}

    

\end{axis}
    
\end{tikzpicture}
\label{fig:experiment_checkout_undo}
\end{subfigure}

\caption{\system's normalized query execution times (w.r.t. Pandas) on 5 TPC-DS queries versus alternative dataframes. Like on the TPC-H queries (\cref{fig:exp_e2e}), \system is up to $1.60\times$ faster than the next best alternative on UDF-heavy queries (e.g., Q7) and scan-heavy joins (e.g., Q96), but falls short on high-cardinality aggregation (e.g., Q3).}
\label{fig:exp_tpcds}
\end{figure}
\paragraph{Performance on Complex Queries} We benchmark the performance of \system versus existing dataframe libraries on 5 complex TPC-DS queries in \cref{fig:exp_tpcds}. \system outperforms \pandas and \modin on all 5 selected queries, but is outperformed by \polars on 2 queries. We observe consistent trends on the TPC-DS queries when compared to the TPC-H queries: \system performs well (1.60$\times$ faster than \polars) on Q7's composite string filtering and Q96's multi-table join aggregation, but underperforms (3.85$\times$ slower than \polars) on Q6, which contains multiple high-cardinality joins on customer keys (\cref{sec:exp_operator}).




\subsection{\systembf Scales Linearly with Data Size}
\label{sec:exp_large}

This section evaluates \system's scalability with varied data sizes. We vary the TPC-H dataset scale from 1GB to 100GB, then measure \system's query execution time versus dataset scale on select TPC-H queries, comparing against existing dataframes. 

We report results in \cref{fig:exp_scalability}. \system demonstrates efficient, near-linear scalability versus dataset scale for all core relational operations like existing parallelized dataframe implementations: For the UDF-heavy Q13 (\cref{fig:mojo_udf}), \system's runtime increases by $11.3\times$ from 1GB to 10GB, which matches the $9.7\times$ and $11.0\times$ scaling exhibited by Polars and Modin, respectively.
For the join and group-by aggregation heavy Q9, \system similarly exhibits near-linear scaling (12.5$\times$) like Modin (12.4$\times$) and Polars (14.7$\times$). In contrast, Pandas shows slightly higher super-linear scaling (13.2$\times$), which degrades further at the 100GB scale (155$\times$ vs. \system's 76$\times$) as it defaults to larger, less compute-efficient datatypes (e.g., INT64~\cite{pandasgroupbybad}) for factorization on higher-cardinality columns (\cref{sec:operation_join}).

\subsection{\systembf Scales to Multiple Cores}
\label{sec:exp_cores}

This section evaluates \system's scalability with different numbers of cores used to perform relational operations. We vary the core number from 2 to 8, then measure \system's query execution time versus core number on select TPC-H queries, comparing against existing dataframe implementations. 

We report results in \cref{fig:exp_cores}. Typical of parallelized dataframe implementations, \system is capable of leveraging multiple cores to achieve query speedups, achieving 1.17$\times$ and 1.34$\times$ speedup on Q9 and Q13, respectively, when increasing the core count from 2 to 8. 

\paragraph{Limitation: lack of fine-grained concurrency control} We observe that \system's speedup is less evident versus Polars and Modin, which achieve $2.18\times$ and $2.29\times$ speedup from 2 to 8 cores on Q9, respectively. This is because Mojo's tools for achieving fine-grained control over thread management and task granularity are still under development~\cite{whymojo}: Mojo currently does not provide control over the number of cores to use per operation (Unlike C++'s \texttt{omp parallel}~\cite{openmp}), and only allows specification of the core limit for an entire program~\cite{mojoparallelize}. This results in significant parallelization overhead in less parallelizable operations and under-utilization of available cores. In contrast, Modin and Polars have access to mature parallel execution frameworks (Ray~\cite{moritz2018ray} and Rayon~\cite{rayon}) in their respective programming languages.

\begin{figure}[t]
\usetikzlibrary{patterns}

\begin{subfigure}[b]{\linewidth}
\centering
    \begin{minted}[fontsize=\scriptsize]{sql}
...
from
customer left outer join orders on (
  c_custkey = o_custkey
  and o_comment not like '%special%requests%'
)
...
    \end{minted}
\caption{Original SQL Operator}
\end{subfigure}
\begin{subfigure}[b]{\linewidth}
\centering
\vspace{2mm}
    \begin{minted}[fontsize=\scriptsize]{sql}
not_string_exists_before(df_orders, o_comment, "special", "requests")
    \end{minted}
\caption{\system Trait Implementation}
\end{subfigure}

\centering
\begin{subfigure}[b]{\linewidth}
\centering
\begin{tikzpicture}

\pgfplotstableread[col sep=comma,]{
name
\criu
\dumpsession
\system \textbf{(Ours)}
}\datatable

\begin{axis}[
    ybar,
    clip=false,
    xlabel style={yshift =-0.5ex},
    width=94mm,
    height=32mm,
    bar width=5mm,
    ymin=0,
    ymax=8,
    ylabel style={yshift = -3ex,align=center},
    axis y line*=none,
    axis x line*=none,
    ytick={0, 2, 4, 6, 8},
    yticklabels={0, 2, 4, 6, 8},
    xtick={1, 2, 3, 4},
    xtick style={draw=none},
    xticklabels={\pandas, \modin, \polars,\system},
    x tick label style={yshift = 1ex},
    xmin=0.5,
    xmax=4.5,
    ymajorgrids,
    tick label style={font=\scriptsize, align=center},
    legend style={
        font=\footnotesize,
        legend columns=4,
        at={(-0.15, 1.1)}, anchor=south west
    },
    label style={font=\footnotesize},
    ylabel={Runtime (s)},
    area legend,
    bar shift=0pt 
]
    \addplot[black,fill=PandasColor] coordinates {(1,6.31)};
    \addplot[black,fill=ModinColor]  coordinates {(2,2.855)};
    \addplot[black,fill=PolarsColor] coordinates {(3,4.165)};
    \addplot[black,fill=OursColor]   coordinates {(4,0.51)};
\end{axis}

\end{tikzpicture}
\caption{Operator Runtime}
\end{subfigure}
\caption{String filtering UDF on the \texttt{o\_comment} column in TPC-H Q13. \system applies this UDF with its stateless trait-based approach $5.60\times$ faster than the next best method.
}
\label{fig:mojo_udf}
\end{figure}

\begin{figure}[t]
\begin{subfigure}[b]{\linewidth}
\centering
    \begin{minted}[fontsize=\scriptsize]{sql}
...
GROUP BY l_orderkey, o_orderdate, o_shippriority
...
    \end{minted}
\caption{Original SQL Operator}
\end{subfigure}
\hfill
\centering
\begin{subfigure}[b]{\linewidth}
\centering
\begin{tikzpicture}

\pgfplotstableread[col sep=comma,]{
name
\criu
\dumpsession
\system \textbf{(Ours)}
}\datatable

\begin{axis}[
    ybar,
    clip=false,
    xlabel style={yshift =-0.5ex},
    width=94mm,
    height=32mm,
    bar width=5mm,
    ymin=0.001,
    ymax=1,
    ylabel style={yshift = -3ex,align=center},
    axis y line*=none,
    axis x line*=none,
    ytick={0.001, 0.01, 0.1, 1},
    yticklabels={0.001, 0.01, 0.1, 1},
    ymode=log,
    log origin=infty,
    xtick={1, 2, 3, 4, 5},
    xtick style ={draw=none},
    xticklabels={\pandas, \modin, \polars, \system, PandasMojo},
    x tick label style={yshift = 1ex},
    xmin=0.5,
    xmax = 5.5,
    ymajorgrids,
    tick label style={font=\scriptsize, align=center},
    legend style={
        font=\footnotesize,
        /tikz/every even column/.append style={column sep=0.5cm},
        legend columns = 4,
        at={(-0.15, 1.1)}, anchor=south west
    },
    label style={font=\footnotesize},
    ylabel={Runtime (s)},
    area legend,
        bar shift=0pt 
    ]
    \addplot[black,fill=PandasColor]coordinates {(1,0.09)};

    \addplot[black,fill=ModinColor]coordinates{(2, 0.741)};

    \addplot[black,fill=PolarsColor]coordinates{(3, 0.011)};

      \addplot[black,fill=OursColor]coordinates{(4, 0.026)};
    \addplot[black,fill=white, postaction={
        pattern=crosshatch
    }]coordinates{(5, 1.04)};

\end{axis}
    
    
\end{tikzpicture}
\caption{Operator Runtime}
\end{subfigure}
\caption{Three-column group-by in TPC-H Q3 (left). \system's group-by (\cref{alg:mojo_groupby}) is $3.5\times$ faster than Pandas' group-by (\cref{alg:pandas_groupby}) on the 10G scale dataset. A direct translation of Pandas' approach to Mojo works poorly (right).
}
\label{fig:mojo_groupby_bench}
\end{figure}
\begin{figure}[t]
\usetikzlibrary{patterns}
\centering
\begin{subfigure}[b]{\linewidth}
    \begin{minted}[fontsize=\scriptsize]{sql}
...
from customer, orders where c_custkey=o_custkey
...
    \end{minted}
\caption{Original SQL Operator}
\end{subfigure}
\hfill
\centering
\begin{subfigure}[b]{\linewidth}
\centering
\begin{tikzpicture}

\pgfplotstableread[col sep=comma,]{
name
\pandas
\system
}\datatable

\begin{axis}[
    ybar,
    clip=false,
    xlabel style={yshift =-0.5ex},
    width=94mm,
    height=32mm,
    bar width=5mm,
    ymin=0.001,
    ymax=1,
    ylabel style={yshift = -3ex,align=center},
    axis y line*=none,
    axis x line*=none,
    ytick={0.001, 0.01, 0.1, 1},
    yticklabels={0.001, 0.01, 0.1, 1},
    ymode=log,
    log origin=infty,
    xtick={1, 2, 3, 4, 5},
    xtick style ={draw=none},
    xticklabels={\pandas, \modin, \polars, \system, SortMerge},
    x tick label style={yshift = 1ex},
    xmin=0.5,
    xmax = 5.5,
    ymajorgrids,
    tick label style={font=\scriptsize, align=center},
    legend style={
        font=\footnotesize,
        /tikz/every even column/.append style={column sep=0.5cm},
        legend columns = 4,
        at={(-0.15, 1.1)}, anchor=south west
    },
    label style={font=\footnotesize},
    ylabel={Runtime (s)},
    area legend,
        bar shift=0pt 
    ]
    \addplot[black,fill=PandasColor]coordinates {(1,0.15)};

    \addplot[black,fill=ModinColor]coordinates{(2, 0.76)};

    \addplot[black,fill=PolarsColor]coordinates{(3, 0.03)};

      \addplot[black,fill=OursColor]coordinates{(4, 0.10)};
    \addplot[black,fill=white, postaction={
        pattern=crosshatch
    }]coordinates{(5, 1.41)};

\end{axis}
    
    
\end{tikzpicture}
\caption{Operator Runtime}
\end{subfigure}
\caption{Joining on unordered join columns in TPC-H Q3 (left). \system adopts Pandas' hash join into Mojo for faster ($1.50\times$ on 10G scale), optimized joins. Specialized alternatives such as sort-merge join underperform (right).
}
\label{fig:mojo_join}
\end{figure}



\subsection{Relational Operator Performance}
\label{sec:exp_operator}
This section breaks down \system's per-operator performance. We isolate the execution time of filter, group-by aggregation, and join within complex TPC-H queries to validate our architectural decisions. We use the 10GB scale TPC-H dataset and compare against existing dataframe and algorithm implementations.

We report results in \cref{fig:mojo_udf}, \cref{fig:mojo_groupby_bench}, and \cref{fig:mojo_join} for filter, group-by aggregation, and join, respectively. 
\system's trait-based approach allows it to declare stateless lambda functions for filtering in a parallelized, vectorized manner (\cref{sec:operations_filter}), enabling it to perform the regex matching operation in TPC-H Q13 5.60$\times$ faster than the next best alternative (\modin). On the other hand, all baseline dataframe implementations cannot take advantage of the applied UDF being stateless and instead apply it across rows sequentially, resulting in the filter application accounting for up to 91.5\% of the query runtime (\pandas) versus \system's 29.1\%.

\paragraph{Limitation: Unoptimized Dictionary in Mojo}
Unfortunately, Mojo's current dictionary implementation, which \system relies on for group-by aggregation (\cref{sec:operations_groupby}) and joins (\cref{sec:operation_join}), is unoptimized for handling large numbers of keys (i.e., distinct elements). One potential issue is open addressing with quadratic probing~\cite{mojobaddict}. This is observed in TPC-H Q3's three-column group-by and join operations, which are respectively performed on the high-cardinality columns \texttt{orderkey} and \texttt{custkey}. Hence, while \system's group-by and join performance outperforms \pandas (3.46$\times$ and 1.50$\times$, respectively) and \modin, it is outperformed by \polars (2.36$\times$ and 3.31$\times$, respectively), which uses the Rust dictionary's SIMD-assisted lookup~\cite{rustdictionary}. However, the Mojo community is actively working on improving Mojo's dictionary implementation~\cite{mojodict}, hence we consider this to not be a limitation inherent to Mojo or \system (\cref{sec:discussion}).

\paragraph{Relational Operator Implementation}
We further verify \system's operator implementation versus alternative native implementations. For group-by, while \system's transposed approach (\cref{sec:operations_groupby}) incurs transposition overhead, a direct translation of \pandas' column-order incremental approach (PandasMojo, \cref{fig:mojo_groupby_bench}) requires maintaining mutable keys, which triggers costly copying even on lookup. As a result, \system's tuple-based composite key computation is 40.0$\times$ faster. For joins, we explored specialized algorithms like sort-merge join~\cite{graefe1994sort} (SortMerge, \cref{fig:mojo_join}); however, naively performing sort-merge on unordered columns incurs heavy performance penalties (14.1$\times$ slower) even with Mojo's vectorized sorting~\cite{mojosort}. Hence, we defer incorporating these algorithms and their dynamic selection (e.g., based on sorted status) to future work.

\subsection{Microbenchmark: Compilation}
\label{sec:exp_compilation}

This section studies the overhead of Mojo's JIT compilation for using \system. We vary the TPC-H dataset scales, number of cores, query structure, and perform JIT compilation for query execution with \system instead of using ahead-of-time compiled code. We measure and compare the time taken for compilation and query compute during \system's end-to-end query execution.

We report results in \cref{fig:exp_compile}. \system's JIT compilation time remains largely constant regardless of the query workload, number of cores, and dataset size, being on average 2.3 seconds with only up to $3\%$ variation across runs.\footnote{This time is equivalent to the time for \system's ahead-of-time compilation, as both leverage the same underlying MLIR/LLVM based compiler infrastructure.} This factor-agnostic JIT compilation time is notably lightweight versus the query compute times of the TPC-H queries on larger dataset scales, contributing to only $10.4\%$ of the end-to-end query execution time (Q21, 10GB).


\subsection{Microbenchmark: Data Loading}
\label{sec:exp_dataload}

This section studies \system's data load speed. To evaluate the efficiency of Mojo's native I/O facilities, we measure the time to ingest columns from raw bytes, comparing against standard CSV loading in existing dataframes. We use the 10GB dataset scale relevant to select queries and load from SSD into \system (i.e., in-memory).

We report results in \cref{fig:exp_dataload}. \system leverages a custom data loader, enabling it to load the 3 required integer columns (\texttt{partkey, suppkey, supplycost}) from TPC-H Q2's \texttt{Partsupp} table. Even against baselines configured to load only necessary columns (via \texttt{usecols}), \system sees a 53.7$\times$ speedup over \modin. Notably, it outperforms \polars by 13.8$\times$ on \texttt{Partsupp} (Q2) and 2.80$\times$ on \texttt{Lineitem} (Q19). Though our data loader uses binary format, it isolates the performance of \system's tensor materialization from the CPU-bound text parsing of CSV. This result demonstrates \system's ability to effectively utilize memory bandwidth, providing insight into implementing a high-performance data loader in the Mojo ecosystem in the future.




\paragraph{Limitation - Lack of Mojo-Native String Tensor}
\system's data loading performance is currently limited by the lack of an efficient Mojo-Native string loader for reading high-cardinality non-numeric columns (\cref{sec:framework}), such as the \texttt{comment} column of TPC-H Q13's \texttt{Orders} table. While Mojo (and \system) has to create individual string objects for each string in the column, \polars utilizes Arrow's \texttt{largestring} format~\cite{largestring} which packs all strings into one compact bytearray, enabling both faster reading times (1.40$\times$ faster on \texttt{Orders}) and smaller in-memory size (\cref{sec:exp_datasize}). Hence, developing a Mojo-native, high-performance string tensor format remains critical future work.

\subsection{Microbenchmark: In-Memory Data Size}
\label{sec:exp_datasize}
\begin{table}[t]
\centering
\caption{\system's in-memory table sizes (GB) on the 10GB scale TPC-H dataset versus baselines.}
\footnotesize
\begin{tabular}{|l|c|c|c|}
\toprule
 \diagbox{Method}{TPC-H Table} & \texttt{partsupp} & \texttt{lineitem} & \texttt{orders}\\
  \midrule
\pandas/\modin & 1.764 & 36.158 & 6.118\\
 \midrule
\polars & 1.372 & 12.651 & 2.445\\
 \midrule
\textbf{\system (Ours)} & 1.406 & 13.489 & 3.395 \\
 \midrule
\cellcolor{gray!20}On-disk CSV & \cellcolor{gray!20}1.171 & \cellcolor{gray!20}7.593 & \cellcolor{gray!20}1.708 \\\bottomrule
\end{tabular}

\label{tbl:experiment_datasize}
\end{table}

\begin{table}[t]
\centering
\caption{\system's in-memory column sizes (GB) on the 10GB scale TPC-H dataset's \texttt{lineitem} table versus baselines.}
\footnotesize
\addtolength{\tabcolsep}{-3pt} 
\begin{tabular}{|l|c|c|c|c|}
\toprule
\multirow{2}{*}{\diagbox{Method}{Column}} & \texttt{orderkey} & \texttt{quantity} & \texttt{returnflag} & \texttt{comment}\\
& \texttt{(INT64)} & \texttt{(DOUBLE)} & \texttt{(CHAR)} & \texttt{(STRING)}\\
  \midrule
\pandas/\modin & 0.479 & 0.479 & 3.479 & 5.008 \\
 \midrule
\polars & 0.479 & 0.479 & 0.539 & 2.069\\
 \midrule
\textbf{\system} & 0.480 & 0.480 & 0.480 & 3.269 \\
\bottomrule
\end{tabular}
\addtolength{\tabcolsep}{3pt} 
\label{tbl:experiment_columnsize}
\end{table}

This section studies \system's in-memory datasize. We load various TPC-H tables (10GB scale) with \system and existing dataframe implementations, and compare the resulting memory sizes with the tables' on-disk sizes.

We report results in \cref{tbl:experiment_datasize}. All dataframe implementations show memory expansion relative to the on-disk size due to various data structure metadata. \system's in-memory table size is up to 1.98$\times$ the on-disk size (\texttt{Orders}). In comparison, \system is significantly more efficient than \pandas (achieving up to 1.80$\times$ space savings) but remains 1.39$\times$ larger than \polars.

\Cref{tbl:experiment_columnsize} breaks down the footprint by inspecting columns of different datatypes in the \texttt{lineitem} table. \system efficiently holds the numeric \texttt{orderkey} and \texttt{quantity} columns with its Mojo-native tensors, like all existing dataframe implementations which use arrays. A significant part of \system's memory size overhead comes from the storage of unmapped, high-cardinality string columns (\cref{sec:framework}): \system uses $1.58\times$ more memory to store the \texttt{comment} column versus \polars, which uses the efficient \texttt{largestring} bytearray format (\cref{sec:exp_dataload}). Unfortunately, there is currently no native alternative in Mojo. \system instead uses individual string objects to store unmapped columns, with each string incurring 20 bytes of overhead~\cite{mojostringlayout}. However, this is still significantly less than the overhead of \pandas, which uses Python strings to store string columns, each containing 49 bytes of metadata~\cite{pythonstringlayout}.

Finally, \system achieves significant savings on the low-cardinality \texttt{returnflag} column; the combined size of the char-to-integer mapping and the tensor is 1.12$\times$ smaller than that of \polars, which stores the characters as is with \texttt{largestring}.


\begin{figure}[t]
\usetikzlibrary{patterns}
\begin{subfigure}[b]{0.48\linewidth}
\centering
\begin{tikzpicture}

\pgfplotstableread{
Label Read Compile Compute
2 12.264386038001248 2.3 6.375627871
4 12.355067460997816 2.3 5.644276198
8 12.348864740000863 2.3 5.447658624
}\testdata

\begin{axis}[
    ybar stacked,
    clip=false,
    xlabel style={yshift = 1.5ex},
    width=45mm,
    height=32mm,
    bar width=2mm,
    xmin=-0.5,
    xmax=2.5,
    ylabel style={yshift=-4ex},
    axis y line*=none,
    axis x line*=none,
    xtick={0, 1, 2},
    xticklabels from table={\testdata}{Label},
    ytick={0,2,4,6,8,10},
    yticklabels={0,2,4,6,8,10},
    y tick label style={yshift = 0ex},
    ymin=0,
    ymax =10,
    ymajorgrids,
    tick label style={font=\footnotesize},
    x tick label style={yshift=0.5ex},
    legend style={
        font=\footnotesize,
        /tikz/every even column/.append style={column sep=0cm},
        legend columns = 4,
        at={(0.6, 1.15)}, anchor=south west
    },
    label style={font=\footnotesize},
    xlabel={Num. cores},
    ylabel={Time (s)},
    area legend
    ]

    \addplot [fill=OursColor] table [y=Compile, meta=Label, x expr=\coordindex] {\testdata};
    \addplot [fill=PandasColor] table [y=Compute, meta=Label, x expr=\coordindex] {\testdata};
    \addlegendentry{Compile}
    \addlegendentry{Compute}

\end{axis}
    
    
\end{tikzpicture}
\caption{Q9}
\end{subfigure}
\begin{subfigure}[b]{0.48\linewidth}
\centering
\begin{tikzpicture}

\pgfplotstableread{
Label Read Compile Compute
1GB 1.1711770580004668 2.3 1.329673669
3GB 1.1784095910006727 2.3 4.476284373
10GB 1.178049091999128 2.3 20.125163353
}\testdata

\begin{axis}[
 ybar stacked,
    clip=false,
    xlabel style={yshift = 1.5ex},
    width=45mm,
    height=32mm,
    bar width=2mm,
    xmin=-0.5,
    xmax=2.5,
    ylabel style={yshift=-4ex},
    axis y line*=none,
    axis x line*=none,
    xtick={0, 1, 2},
    xticklabels from table={\testdata}{Label},
    ytick={0,10,20,30},
    yticklabels={0,10,20,30},
    y tick label style={yshift = 0ex},
    ymin=0,
    ymax =30,
    ymajorgrids,
    tick label style={font=\footnotesize},
    x tick label style={yshift=0.5ex},
    legend style={
        font=\footnotesize,
        /tikz/every even column/.append style={column sep=0cm},
        inner ysep=0.1pt,
        legend columns = 4,
        at={(0.6, 1.05)}, anchor=south west
    },
    label style={font=\footnotesize},
    xlabel={Dataset Scale},
    ylabel={Time (s)},
    area legend
    ]

    \addplot [fill=OursColor] table [y=Compile, meta=Label, x expr=\coordindex] {\testdata};
    \addplot [fill=PandasColor] table [y=Compute, meta=Label, x expr=\coordindex] {\testdata};

\end{axis}
    
    
\end{tikzpicture}
\caption{Q21}
\end{subfigure}

\caption{Breakdown of \systembf's JIT compilation and query compute times for end-to-end query execution versus query, num. cores (left) and dataset scale (right). Compilation time is factor-agnostic, and negligible versus compute times.}
\label{fig:exp_compile}
\end{figure}

\begin{figure}[t]
\usetikzlibrary{patterns}
\centering
\begin{tikzpicture}

\pgfplotstableread[col sep=comma,]{
name
Partsupp (Q2)
Lineitem (Q19)
Orders (Q13)

Customer
}\datatable

\begin{axis}[
    ybar,
    clip=false,
    xtick={1, 2, 3},
    xticklabels from table={\datatable}{name},
                 x tick label style={anchor=center, yshift = 0ex, font=\footnotesize},
    xtick style ={draw=none},
    xlabel style={yshift = 2.5ex, font=\footnotesize},
    ylabel style={yshift = -3.5ex, font=\footnotesize},
    width=85mm,
    height=32mm,
    bar width=2mm,
    ymin=0,
    ymax=30,
    axis y line*=none,
    axis x line*=none,
    ytick={0, 10, 20, 30},
    yticklabels={0, 10, 20, 30},
    xmin=0.5,
    xmax = 3.5,
    ymajorgrids,
    tick label style={font=\footnotesize},
    legend style={
        font=\footnotesize,
        /tikz/every even column/.append style={column sep=0.2cm},
        legend columns = 4,
        at={(0.46,1.4)},
        anchor=center,
                legend image post style={xscale=0.6},
        /tikz/every even column/.append style={column sep=4pt}
    },
    ylabel={Runtime (s)},
    area legend
    ]

\addplot[black, fill=PandasColor]
table[x=x,y=y] {
x y
1 3.001573653004016
2 26.731481437003822
3 9.177387299001566

};
\addlegendentry{\pandas}

\addplot[black, fill=ModinColor]
table[x=x,y=y] {
x y
1 1.9671021009999095
2 13.464148623002984
3 3.5984671559999697

};
\addlegendentry{\modin}

\addplot[black, fill=PolarsColor]
table[x=x,y=y] {
x y
1 0.5112237440043828
2 4.915656776000105
3 0.852384986996185
};
\addlegendentry{\polars}

\addplot[black, fill=OursColor]
table[x=x,y=y] {
x y
1 0.037
2 1.753
3 1.219
};
\addlegendentry{\system \textbf{(Ours)}}

    

    

\end{axis}

    
\end{tikzpicture}

\caption{Data loading times for TPC-H tables (10G scale) with \systembf versus alternative dataframes. \system can load only required columns (projection pushdown). This shows efficient tensor materialization and peak throughput; numeric data (\texttt{partsupp}) is loaded up to 13.8$\times$ faster than alternatives.}
\label{fig:exp_dataload}
\end{figure} 

\section{Related Work}
\label{sec:related_work}
\paragraph{Existing Mojo Libraries}
There currently exists a large variety of libraries in Mojo~\cite{awesomemojo}:  (1) libraries for AI pipelines such as machine learning algorithms~\cite{mojostats}, StableDiffusion~\cite{mojostablediffusion}, and LLMs~\cite{mojollama}, (2) domain-specific libraries such as audio processing~\cite{mojoaudio}, quantum computing~\cite{mojoquantum}, and bioinformatics~\cite{mojobio}, (3) libraries that extend Mojo with additional data structures such as arrays~\cite{mojoarray}, trees~\cite{mojotree}, dictionaries~\cite{mojodictionary}, and queues~\cite{mojoqueue}, and (4) libraries for system programming such as networking~\cite{mojonetwork} and logging~\cite{mojolog}.
We add \system---dataframe library for Mojo on which relational operations can be performed to the Mojo ecosystem.

\paragraph{GPU-based Analytics}
Accelerating analytical tasks such as performing relational operations by using GPU acceleration is a well-studied problem~\cite{treinen2008description, nishino2017cupy, yogatama2023accelerating, yogatama2024scaling, crystal, ocsa2019sql, gao2021scaling, yogatama2022orchestrating}.
cuDF~\cite{treinen2008description} and cuPy~\cite{nishino2017cupy} are CPU-GPU dataframe libraries which allow users to specify which of CPU or GPU to use for data placement and/or computations.
BlazingSQL~\cite{ocsa2019sql} and Crystal~\cite{crystal} are GPU databases that supports executing SQL queries with GPUs.
Gao et. al. proposes a method for speeding up joins with multiple GPUs~\cite{gao2021scaling}.
There are also works aimed at optimizing data placement~\cite{yogatama2022orchestrating, yogatama2024scaling} and performing JIT~\cite{yogatama2023accelerating} for GPU computations.
Recently, Tensor Query Processing (TQP~\cite{he2022query}) has been proposed for running queries with the PyTorch library~\cite{pytorch} on GPU, like \system, integrates string operations into tensor computations via custom techniques (string padding). However, to the best of our knowledge, we have not found an open-source version of TQP for fair comparison.
We design the data structure of \system, our Mojo-based dataframe, to be mainly tensor-based to natively support GPU acceleration (\cref{sec:framework}).

\paragraph{Query Optimization for Dataframe Workloads} Recently, the \polars library introduced the LazyFrame class for lazy query execution~\cite{pllazyframe}, which can perform common DBMS optimizations such as predicate pushdown~\cite{jeong2025upp} for faster execution. 
While \system currently does not support lazy execution and we disable \polars' lazy execution for fair comparison on individual operator performance (\cref{sec:exp_setup}), incorporating lazy execution into \system can be valuable future work.











\paragraph{Just-In-Time Compilation for Data Science}
JIT compilation has been extensively explored for speeding up data science code in interpreted languages such as Python and R~\cite{lavrijsen2016high, lam2015numba, graalpy, pyston, rtorch, rcompiler}.
Numba~\cite{lam2015numba} uses the LLVM compiler to optimize NumPy arrays and functions by applying threading and SIMD.
PyPy~\cite{lavrijsen2016high} is an alternative Python interpreter featuring a tracing JIT compiler that performs established optimizations such as hot loop tracing~\cite{bolz2009tracing}.
R contains a native JIT compiler package~\cite{rcompiler} with adjustable JIT levels controlling which code structures (e.g., closures, control flows) are compiled for different compilation time-runtime trade-offs, and the Torch library~\cite{rtorch} for accelerating array operations for machine learning. 
\system implements relational operations (e.g., join, \cref{sec:operation_join}) in ways that take advantage of Mojo's JIT compilation.

\paragraph{Code-Generating Database Systems}
Prior database research has extensively studied compilation and JIT-based execution strategies. Systems such as HyPer and Umbra introduced dynamic LLVM IR generation to optimize relational algebra query pipelines ~\cite{kemper2011hyper, Neumann2014CompilingDQ}. LegoBase~\cite{klonatos2014legobase} uses Scala to generate C-like low-level code. Additionally, frameworks like the Tensor Data Platform~\cite{tensordata} and "Share the Tensor Tea"~\cite{sharetensor} have explored embedding tensor primitives directly into relational query execution. \system differs from these existing approaches in several aspects: (1) First, imperative dataframe operations are directly compiled through Mojo’s native MLIR pipeline, rather than utilizing traditional SQL-based relational algebra frameworks;
this approach provides greater flexibility for exploratory data analysis and iterative development. Dataframes natively support dynamic schema evolution, heterogeneous data types, without requiring predefined table schemas or normalization constraints. Unlike traditional relational models that prioritize data integrity and transactional consistency, dataframes are great for ad-hoc transformations and handling data that is not perfectly structured common in modern analytics workflows. (2) Second, \system's architecture differs from those that support tensor query execution, specifically in the data storage model. \system uses a tensor-native design, where each column in the dataframe is backed by a tensor data structure. This contrasts with TDP and similar systems, which begin with a relational schema and use tensor operations only after the query planning stage.

\paragraph{Systems for Speeding up Data Science Coding} There exist works for speeding up the coding process for building data science pipelines~\cite{wang2022diff, li2023elasticnotebook, li2024kishu, li2023edassistant, lee2021lux, suris2023vipergpt, eghbali2024hallucinator, bauerle2022symphony, wu2020b2}.
Code completion tools recommend next lines of code for the user via either rule-based~\cite{li2023edassistant, lee2021lux} or LLM-based~\cite{suris2023vipergpt, eghbali2024hallucinator} predictions.
Checkpointing tools such as Diff-in-the-loop~\cite{wang2022diff}, ElasticNotebook~\cite{li2023elasticnotebook, li2024demonstration}, Kishu~\cite{li2024kishu, chockchowwat2023transactional, li2025demo}, and Chipmink~\cite{chockchowwat2025chipmink} can be used to save states of data science pipelines for returning to later, facilitating more efficient code iteration.
Symphony~\cite{bauerle2022symphony} and B2~\cite{wu2020b2} adopt a non-coding approach and enable point-and-click interactions with ML models and dataframes.
\system enables users to more conveniently write and run (Mojo-native) data science pipeline code in Mojo by eliminating the need to import special-purpose libraries (\cref{sec:background_others}) or alter code based on available hardware (\cref{sec:background_gap}).


\section{Conclusion}

In this paper, we presented \system, the first DataFrame library native to Mojo. 
\system is built on Mojo's high-performance tensor operations and JIT compilation with MLIR for efficiently running relational operations.
We designed a cardinality-aware hybrid representation to efficiently integrate non-numeric columns, which are not natively supported by Mojo's numeric tensors.
Then, for relational operations, we formulate per-operation optimizations that take advantage of the Mojo language's JIT compilation and developed algorithms and optimizations that navigate the language's current constraints regarding mutable state. We show that \system supports all 22 queries in the TPC-H benchmark and 5 TPC-DS queries, which cover a wide range of relational operations, achieving up to $4.60\times$ speedup compared to established dataframe libraries. Our findings validate Mojo's and \system's potential for data engineering, while pinpointing specific bottlenecks regarding in-memory data representations and dictionary operations.


\section{Acknowledgments}

This work was supported in part by the National Science Foundation under Awards \#2440498 and \#2312561.

\section*{AI-Generated Content Acknowledgement}
We acknowledge the use of generative AI tools, specifically OpenAI's ChatGPT, Anthropic's Claude, and Google's Gemini, solely as a writing aid to improve grammar and to assist with formatting. All scientific claims, code, and final text are original. All substantial work and insights remain entirely our own.

\bibliographystyle{IEEEtran}
\bibliography{sample}

\end{document}